\newcommand\lam{\mbox{$\:\lambda\ $}}
\newcommand\lamlam{\mbox{$\:\lambda\lambda $ }}
\newcommand\ha{{H$\alpha$}}

\newcommand\kms{\:\rm{km\,s^{-1}}}

\newcommand\VEL{\:{\rm km\:s^{-1}}}

\newcommand\OiL{[\ion{O}{1}] $\lambda 6300$}

\newcommand\SiiLL{[\ion{S}{2}] $\lambda\lambda 6716, 6731$}
\newcommand\NiiLL{[\ion{N}{2}] $\lambda\lambda 6548, 6583$}

\newcommand\FeiiL{[\ion{Fe}{2}] 1.644 $\mu$m}

\newcommand\sii{[\ion{S}{2}]}
\newcommand\nii{[\ion{N}{2}]}
\newcommand\oi{[\ion{O}{1}]}

\newcommand\oiii{[\ion{O}{3}]}

\newcommand\hii{\ion{H}{2}}

\newcommand\gal{NGC\,4449}

 \newcommand{\SFR}{\mbox{$\:M_{\sun}\;{\rm yr}^{-1}$}}  

\documentclass[]{modaastex631}

\usepackage[utf8]{inputenc}
\usepackage{natbib}
\usepackage{comment}
\usepackage{todonotes}
\usepackage{graphicx}
\usepackage{lineno}
\usepackage{comment}
\usepackage{amssymb}
\turnoffediting

\begin{document}

\flushleft{Accepted for publication in {\em The Astrophysical Journal
\bigskip
}

\title{Supernova Remnants in the Irregular Galaxy NGC\,4449}

\shorttitle{SNRs in Irregular Galaxy NGC\,4449}

\correspondingauthor{P. Frank Winkler}
\email{winkler@middlebury.edu}

\author[0000-0001-6311-277X]{P. Frank Winkler}
\affil{Department of Physics, Middlebury College, Middlebury, VT, 05753; 
winkler@middlebury.edu}
    \affil{Visiting astronomer, Kitt Peak National Observatory, National Optical Astronomy Observatory, which is operated by the Association of Universities for Research in Astronomy (AURA) under a cooperative agreement with the National Science Foundation.}

\author[0000-0002-4134-864X]{Knox S. Long}
\affil{Visiting astronomer, Kitt Peak National Observatory, National Optical Astronomy Observatory, which is operated by the Association of Universities for Research in Astronomy (AURA) under a cooperative agreement with the National Science Foundation.}
\affil{Space Telescope Science Institute,
3700 San Martin Drive,
Baltimore MD 21218, USA; long@stsci.edu}
\affil{Eureka Scientific, Inc.
2452 Delmer Street, Suite 100,
Oakland, CA 94602-3017}

\author[0000-0003-2379-6518]{William P. Blair}
\affil{Visiting astronomer, Kitt Peak National Observatory, National Optical Astronomy Observatory, which is operated by the Association of Universities for Research in Astronomy (AURA) under a cooperative agreement with the National Science Foundation.}
\affil{The William H. Miller III Department of Physics and Astronomy, 
Johns Hopkins University, 3400 N. Charles Street, Baltimore, MD, 21218; 
wblair@jhu.edu}


\begin{abstract}

The nearby irregular galaxy \gal\ has a star formation rate of $\sim0.4\SFR$ and  should host of order 70 SNRs younger than 20,000 years, a typical age for SNRs expanding into to an ISM with a density of 1 cm$^{-3}$ to reach the radiative phase. We have carried out an optical imaging and spectroscopic survey in an attempt to identify these SNRs.  This task is challenging because diffuse gas with elevated ratios of \sii:\ha\ is omnipresent in \gal,  {causing confusion when using this common diagnostic for SNRs.  Using narrow-band interference-filter images, we first identified 49 objects that  have  elevated \sii:\ha\  ratios compared to nearby \hii\ regions.  Using Gemini-N and GMOS, we then obtained high-resolution spectra of 30 of these SNR candidates, 25 of which have \sii:\ha\ ratios greater than 0.5. Of these, 15 nebulae are almost certainly SNRs, based on a combination of characteristics:  higher \oi:\ha\ ratios and  broader line widths than observed from \hii\ regions.  The remainder are good candidates as well, but need additional confirmation. Surprisingly, despite having superior imaging and spectroscopic data sets to examine, we are unable to confirm most of the candidates suggested by  \citet{leonidaki13}.  While \gal\ is likely an extreme case because of the high surface brightness and elevated \sii:\ha\ ratio of diffuse gas, it highlights the need for sensitive high-resolution optical spectroscopy, or high spatial resolution radio or X-ray observations that can ensure accurate SNR identifications in external galaxies.}

\end{abstract}


\keywords{galaxies: individual (NGC\,4449) -- galaxies: ISM  -- supernova remnants}

\section{Introduction}

Supernova remnants (SNRs) are the visible evidence of the shock driven by supernovae (SNe) into the surrounding circumstellar and interstellar medium.  Along with the stellar winds arising from SN progenitors, SNRs are responsible for much if not all of the hot gas in normal galaxies and are an important part of the process of reprocessing interstellar gas into the next generation of stars.  The interaction of the blast wave with the ISM results in radiation throughout the electromagnetic spectrum.  In the Galaxy, most SNRs were first identified as bright, extended, non-thermal radio sources \citep[see, e.g.][]{long17}.   Beyond the Local Group, there are examples of SNRs that have been discovered by this method \cite[see, e.g.][]{lacey97}, but the numbers are relatively few due to limited sensitivity and angular resolution.  Instead, most SNRs and SNR candidates in external galaxies have been identified optically, as isolated nebulae with \sii:\ha\ ratios higher than expected from photoionized regions \citep[cf.\ sec.\ 6 of][for further discussion]{long22}. 

The hypothesis that high \sii;\ha\ ratios can effectively  identify SNRs has  both observational and theoretical bases.  First, optical spectra of SNRs in the Galaxy and the Magellanic Clouds show that the optical spectra of SNRs identified by other means typically show   \sii:\ha\  ratios  of greater than 0.4, whereas (bright) \hii\ regions have ratios of order 0.1 or less.  Secondly, theoretical models of \hii\ regions calculated with photoionization codes such as {\tt Cloudy} or {\tt Mappings} \citep{ferland98, allen08} indicate that \hii\ regions should have low \sii:\ha\ ratios because most of the S in \hii\ regions is ionized to S$^{++}$, whereas shock models 
show that fully radiative SNR shocks have an extended region where S$^+$ is the dominant ion, which  leads to a much higher observed \sii:\ha\ line ratios \citep{raymond79, hartigan87, allen08}. 

Based on this hypothesis, we and others have identified a large number of SNRs and SNR candidates in galaxies within and beyond the Local Group. To date, almost all this attention has gone to spirals, including 
M31 \citep{blair81,lee14a}, M33 \citep{lee14b,long18}, 
M81 \citep{matonick97,lee15}, M83 \citep{blair12,winkler17,williams19,long22}, M51 \citep{winkler21}, NGC\,6946 \citep{long19, long20}, NGC 3344 \citep{moumen19}, NGC 4030 \citep{cidfernandes21}, and others \citep{kopsacheili21}.\footnote{See \citet{vucetic15} for a more complete listing of relevant papers prior to 2015.}

By contrast, with the exception of the Magellanic Clouds \citep{mathewson83,mathewson84,mathewson85,filipovic08,maggi16,bozzetto17}, where radio and X-ray observations are sufficiently sensitive to provide confirmation for SNR identifications, little attention has gone to SNRs in irregular galaxies.  Yet many such galaxies have high star-formation rates that should result in a significant population of SNRs.  Irregular galaxies  typically have much lower metallicities, implying conditions quite different from those in most large spiral galaxies. The most significant attempt to identify SNRs in irregular galaxies was carried out by \cite{leonidaki13} who searched for SNRs in three irregular galaxies (NGC\,3077, NGC\,4214, and \gal) using a combination of interference-filter imaging and follow-up spectroscopy.  In the case of \gal, they identified 19 objects as SNRs based on a spectroscopically determined \sii:\ha\ ratio greater than 0.4.  Additionally they listed  39 objects as ``probable SNRs'', based on photometrically-determined \sii:\ha\  line ratios  greater than 0.4, and 13 objects as ``probable candidate SNRs'' with photometrically-determined ratios between 0.3 and 0.4, bringing their total to 71 objects of interest.

\gal\ is an interesting irregular galaxy to study because it is relatively nearby at 4.2 Mpc \citep{karachentsev03}, where 1\arcsec\ corresponds to 20 pc. It has a robust star formation rate of  $0.4\, M_{\sun}\; {\rm yr}^{-1}$ \citep{chyzy11,manna23}, greater than that of the LMC \citep[$0.2\,  M_\sun\;  {\rm yr}^{-1}$;][]{whitney08}.  Reflecting this high SFR, it is the host of a very young SNR with very broad emission lines of oxygen, widely known as \gal-SNR-1 (henceforth simply SNR-1),  that likely resulted from an explosion of $>$\,20  $M_{\sun}$ star about 65 years ago \citep{blair83, milisavljevic08,mezcua13}.  
A few of the sources identified by \citet{leonidaki13} as SNRs or SNR candidates are coincident or nearly coincident with X-ray sources identified in relatively short  {\em Chandra} observations of the galaxy \citep{summers03} and/or radio sources that  \citet{chomiuk09} identified as as probable SNRs, based on their spectral indices. 

In our own past research, we have  concentrated on local spiral galaxies with significant star formation. 
We were also aware that irregular galaxies such as \gal\ have outflows and that the warm ionized gas contained in those outflows  often shows significantly higher \sii:\ha\ ratios and higher surface brightnesses than is typical for the diffuse emission in spiral galaxies.  As a result, one could anticipate that identifying SNRs by the standard \sii:\ha\ technique might be expected to be more difficult. 

Here we report  the results from new emission-line imaging surveys of \gal, plus follow-up multi-object spectroscopy,  with both significantly higher angular resolution and sensitivity than that of \citet{leonidaki13}.  For identification of nebulae as SNRs, we rely on additional diagnostics beyond the traditional \sii:\ha\ ratio:  the strength of the \oi\ lines, which gives another indicator of cooling gas behind SN shocks, and the velocity width of the emission lines, as demonstrated by \citet{chu88} and \citet{points19}. We propose a new list of  49 SNRs and candidates in \gal.
For a variety of reasons, as discussed in the sections below, our list of SNRs is largely disjoint from that of  \citet{leonidaki13}.

The paper is organized as follows: section 2 discusses the imaging and spectroscopic data sets  used in this paper as well as their processing.  Section 3 details our search for candidate SNRs and the spectroscopic analysis.  We discuss our results  in section 4, in the context of   previous work and of other galaxies. Finally, we summarize our results and conclusions in section 5.

\section{Observations and Data Sets \label{sec:observations}}

\subsection{Imaging Observations: WIYN and Gemini-N}

We initially carried out narrow-band imaging observations of \gal\ from the 3.5m WIYN telescope and MiniMosaic imager on Kitt Peak on the nights of 2011 June 26-28 (UT).\footnote{The WIYN Observatory is a joint facility of the NSF's National Optical-Infrared Astronomy Research Laboratory, Indiana University, the University of Wisconsin-Madison, Pennsylvania State University, the University of Missouri, the University of California-Irvine, and Purdue University.}  The so-called ``Minimo'' was mounted at the f/6.3 Nasmyth port and consisted of a pair of $2048\times4096$ SITe chips, with a field  9\farcm 6 square at a scale of 0\farcs 14 pixel$^{-1}$, easily encompassing the entirety of \gal.  We used interference filters that passed lines of \ha, \sii \lamlam6716,6731, and \oiii \lam 5007, plus red and green narrow-band continuum filters so we could subtract the stars and produce pure emission-line images (see Table 1 for details).  Frames in each filter were dithered to facilitate removal of cosmic rays and bad pixels.  The seeing was variable and not particularly good: our final combined images used the best three exposures through each filter and had effective seeing of 1\farcs 2 -- 1\farcs 5 (FWHM).  On the basis of these images, we assembled a preliminary SNR candidate list of over fifty objects in \gal.

In order to obtain a deeper and higher-resolution view of nebulae in \gal, to refine our list of candidate SNRs, and to facilitate the preparation of accurate masks for follow-up MOS spectroscopy, we obtained imaging data from the GMOS instrument on the 8.1\,m Gemini-North telescope through program  GN-2021A-Q-222.  All the data were taken on 2021 Apr 6, and are detailed in Table 1.  
Processing was carried out using standard procedures in the IRAF\footnote{IRAF is distributed by the National Optical Astronomy Observatory, which is operated by the Association of Universities for Research in Astronomy, Inc., under cooperative agreement with the National Science Foundation.} {\tt gemini} package.  The resolution of the combined GMOS images was 0\farcs 6 -- 0\farcs 7, indeed significantly better than in the earlier  WIYN data.

Fig.\ \ref{fig:s2_by_ha} highlights a striking difference between \gal\ and the spiral galaxies we have studied, in the form of a \sii:\ha\ ratio image formed from the continuum-subtracted GMOS images.  To create this image, we first fit and subtracted a constant value to set the  background sky level to zero.  We then set all pixel values below a threshold to a constant value for both the \sii\ and \ha\ images before taking the ratio of the two. Thus the background sky surrounding the galaxy has a constant grey level in the figure: in this case corresponding to a moderate value of \sii:\ha\ = 0.25.

\begin{figure}
\plotone{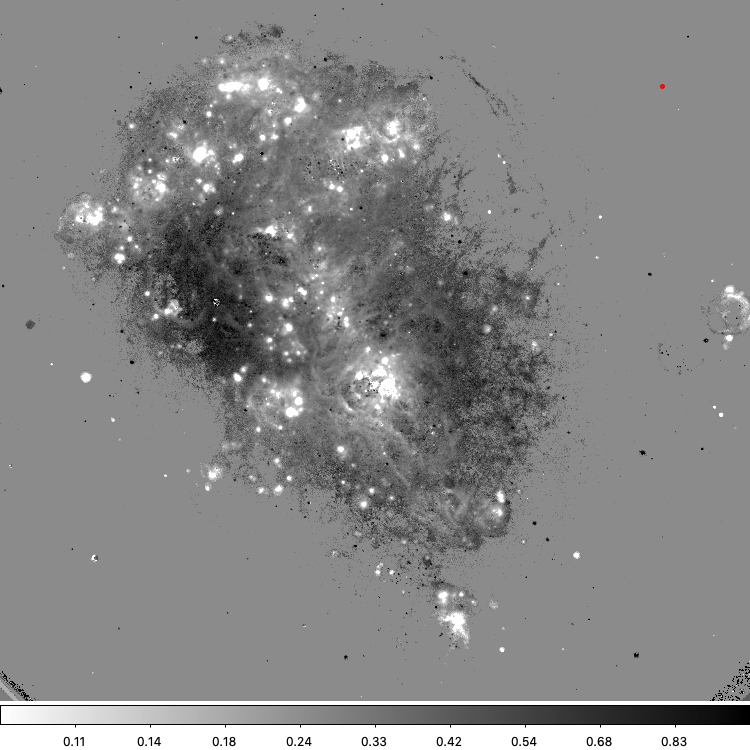}
\caption{Ratio image of \sii:\ha\ for \gal\ derived from GMOS image data.  Low-level values have been clipped for both the \sii\ and \ha\ images before taking the ratio, so that the background sky appears as a constant grey level, at a value corresponding to \sii:\ha = 0.25.  Regions darker than the background have higher ratios and lighter regions have lower ratios.  Bright \hii\ regions show as white, indicating very low ratios, but much of the diffuse emission in \gal\ has an unusually high value, 
\sii:\ha $\gtrsim 0.4$, 
and only the bright photoionized regions have low values for this ratio.  
The field  is 5\arcmin\ square, oriented N up, E left.
\label{fig:s2_by_ha}}
\end{figure}

From this image, it is clear that  most of the diffuse emission throughout \gal\ has a relatively high \sii:\ha\ ratio,  approaching and even exceeding 0.4, the usual criterion for identifying shock-heated nebulae.  This is consistent with the measurement by  \citet{kobulnicky99} of the global \sii:\ha\ ratio for \gal\ as about 0.28.  While we do not have global ratios for the spiral galaxies we have studied, spectra of faint diffuse emission typically shows \sii:\ha\,  $\lesssim 0.15$, characteristic of photoionized material.\footnote{We note that at the lowest surface brightnesses sampled in M83, we encountered much the same situation, where the diffuse ionized gas had elevated values of \sii:\ha, causing confusion in the identification of SNR candidates}; see \cite{long22}.  It is also interesting that the global ratio for \gal\ given by \citet{kobulnicky99} is significantly higher than those for any of the other seven Irr galaxies they observed (0.10 -- 0.19; see their Table 1). As a result, the \sii:\ha\ ratio criterion is not as effective a diagnostic for SNRs in \gal\ as it has been in numerous earlier studies.  This causes significant uncertainty in identification of nebulae that are near this usual dividing line, as discussed in section 3.1.

\subsection{{\em HST} Imaging}   
While we did no {\em HST} imaging of our own for this project,
the Multimission Archive for Space Telescopes (MAST\footnote{\url{https://archive.stsci.edu}.}) contains {\em HST}/ACS imagery of \gal\ in  \ha\ (F658N filter), a combination of \ha\ and \NiiLL\ (F660N),  \oiii\ $\lambda$5007 (F502N), and broadband optical filters \citep{annibali08}. However, there are no images for \SiiLL\ or for \FeiiL, for which WFC3 would have been required.
We downloaded the processed versions of these data from the Hubble Legacy Archive)\footnote{\url{https://hla.stsci.edu}.} for use in this project, although subsequent work to improve the astrometric solutions (using Gaia stars) was required to make accurate comparisons with other data sets. 

The {\em HST}/ACS imagery of course has exquisite spatial resolution (pixel size 0\farcs10), and the bright emission structures and/or small angular size objects are well detected. However, many of the nebulae of interest from our ground-based imaging are extended and have relatively low surface brightness.  Hence, the {\em HST} data have been useful for examining the morphology and physical sizes of some, but by no means all, of the objects of interest.

\subsection{Spectroscopic Observations}
As part of the same Gemini program as the imaging discussed above, we were approved to carry out follow-up spectroscopy: high-resolution MOS red spectra using the B1200 grating for two masks, plus a lower resolution and broader spectral coverage longslit spectrum with the B600 grating of the bright young remnant \gal-SNR-1\footnote{The spectrum of \gal SNR-1 will be reported in a separate paper specific to this object.}.  However, none of these spectra were actually obtained during semester 2021A due to COVID restrictions.  All of the spectral were eventually obtained the following year, through program GN-2022A-Q-130.  The detector was binned $2\times 2$; the spectral range for the B1200 MOS spectra varied with slit mask position, but included wavelengths from well below \OiL\ to well above the \SiiLL\ lines for all objects, with a dispersion  of 0.53 \AA\ per (binned) pixel, and a resolution of $65 \kms$ (FWHM, measured from night-sky lines)\@.  Further details of all the spectroscopic observations are given in Table 2.
 
Through our work on M83 \citep{winkler23}, we had become convinced  that obtaining higher-resolution spectra was important for establishing with high confidence that objects are SNRs.  This becomes especially important for \gal, because of the somewhat compromised nature of the \sii:\ha\ ratio criterion.   The velocity widths for SNRs should be determined largely by the shock velocity and bulk motions of the gas, typically $\gtrsim 100 \kms$ for all but the oldest and most evolved remnants.  This is in contrast to photoionized regions for which we expect much smaller velocity widths, more characteristic of the thermal velocity of the ISM \citep{chu88, points19}, which would essentially be unresolved. In conjunction with a previous program on GMOS-S \citep{winkler23}, we had carried out a series of tests that clearly showed that the B1200 grating with relatively narrow 0\farcs 6 slits gave the optimum combination of resolution and throughput, so we adopted the equivalent combination on GMOS-N for our \gal\ program.  We used the GMOS images described above to design our MOS masks, with slit width 0\farcs 6 for all the objects.    (See section 3.1 for more detail.) 

Immediately before or after  each set of object spectra, we obtained quartz flats and CuAr arcs for calibration.  The spectra were reduced using the {\tt gemini} package in IRAF, including separation into 2-D spectra for each slit.  We then subtracted the local background, stripped out 1-D spectra for each object, and finally did flux calibration based on observations of the spectrophotometric standard star Feige 66, provided by the Gemini Observatory.

\section{Results and Analysis}

\subsection{Identification of Candidate SNRs
\label{sec:image_results}}
  
We identified candidates by conducting a visual search, as follows: we displayed various images, typically the continuum-subtracted \ha\ and \sii, the \sii:\ha\ ratio image,
and a continuum band, to
identify and eliminate possible stellar residuals from consideration. The SAOimage ds9 tool \citep{joye03}
was indispensable for this task.  In the case where
additional imagery is available (e.g., {\em HST} or other multi-wavelength images),
these could be added into the visual search process.

We had conducted an initial search using the WIYN images, and for our final search we used  the more recent and significantly better-seeing images from Gemini GMOS\@.  (Our initial set of objects based on the WIYN data turned out to agree remarkably well with the final GMOS set, though the GMOS images were much better for detailed examination of the candidates, and for diameter measurement.)  Searches were carried out independently by different ones of us, and also independently of the list of candidates suggested by \citet{leonidaki13}.
Since so much diffuse emission with elevated ratio is present, we added the
following aspect to our search strategy: we identified nebulae that had reasonable
integrity to their appearance either in the subtracted images and/or
the \sii:\ha\ ratio image so that they constituted identifiable objects.  Also, in
the case of confused regions of high-ratio background, we attempted
to select objects that were elevated even more than the background.
While clearly we could be incomplete in our candidate list, the addition
of these criteria was intended to improve the likelihood that the selected objects are indeed good candidates.

\begin{figure}
\plotone{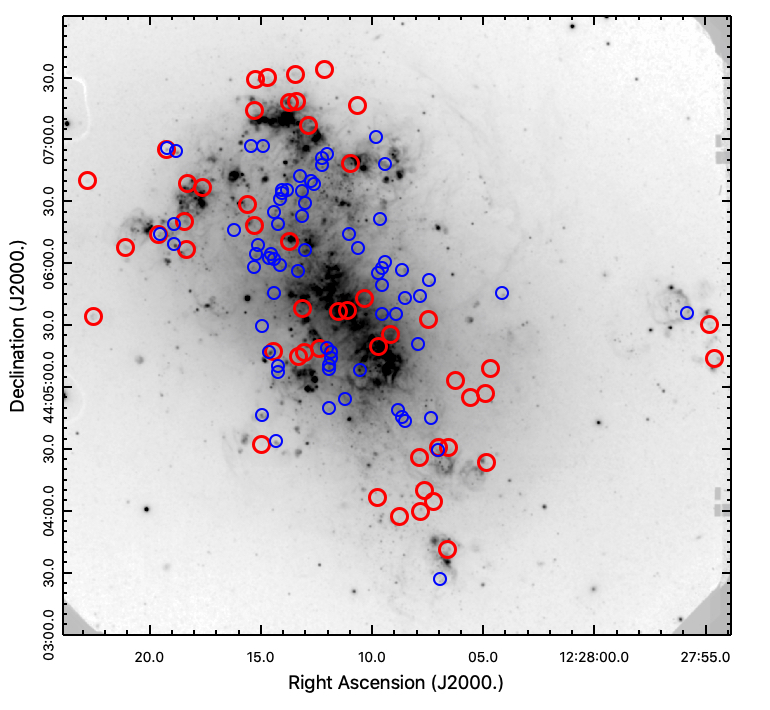}
\caption
{\ha\ image of \gal\ from GMOS, with our newly identified SNRs and candidates marked by red circles.  Also shown are the SNRs and candidates from \citet{leonidaki13}, marked by blue circles.  There are very few coincidences between these two independently selected samples.
\label{fig:index}}
\end{figure}

With the GMOS search complete, we  also made comparisons against the positions
of radio and X-ray point sources tabulated by others
\citep{chomiuk09, summers03, leonidaki10}. Examination of the {\em HST}/ACS
imagery was also useful in this process, and  allowed identification of
sources associated with star clusters and likely X-ray binaries. 
The only coincidences between our candidate objects and {\em Chandra} ones from \citet{summers03} are the well-known SNR-1, the bright, complex nuclear source (W23-22, near coincident with S03-18), and W23-36, coincident with Summers source 24, which they identify as a background AGN.
There are four coincidences between our candidates and radio sources identified by \citet{chomiuk09}, but only three have non-thermal radio spectra characteristic of SNRs:  W23-28 = CW09-19, W23-45 = CW09-26. plus the obvious SNR-1\@.\footnote{In addition, W23-23 is located within 3\arcsec\ of  both CW09-14, listed by them as a SNR, and (more closely) with CW09-15, classified by them as an \hii\ region.}  It is perhaps not surprising to find so few matches, especially in the case of the {\em Chandra} data, which were relatively shallow compared with other nearby galaxy data sets that are available \citep[cf.][]{ long14}.  All the source matches are listed in Table 3 under the column ``Other names."

Our final list contains a total of 49 SNR candidates, which we 
list in Table 3\footnote{We choose not to list the unique and well-known SNR-1 in this Table as we consider it to be in a class by itself.} and show in Fig. \ref{fig:index} as red circles projected onto the GMOS \ha\ image.
Table 3 includes some ancillary data that were derived as we performed
the search.  Size information was determined by setting circular  or elliptical ds9 regions  on
each object identified; where elliptical regions were used,
we list the diameter as the geometric mean of the major and
minor axes of the ellipse.  The {\em HST}/ACS data were used for measuring
object sizes when they were visible, but many of the objects are outside the ACS field or are too low in 
surface brightness to be seen in the available data.
Angular sizes were converted to pc assuming a distance of 4.2 Mpc.
We also include a morphology `Class' assessment: the class A objects 
are either extended objects with evidence for a shell-like morphology or small-diameter objects that stand out against the background.  These are 
more robust in appearance and/or are less confused than the class B objects, which have enhanced \sii:\ha\ ratio relative to the surrounding background, but that are more confused by surrounding nebulosity.  The possible utility of
this criterion is discussed further below, and several examples are shown in Fig.\ \ref{fig:4panels}.

\begin{figure}
\vspace{-6mm}
\gridline{
\fig{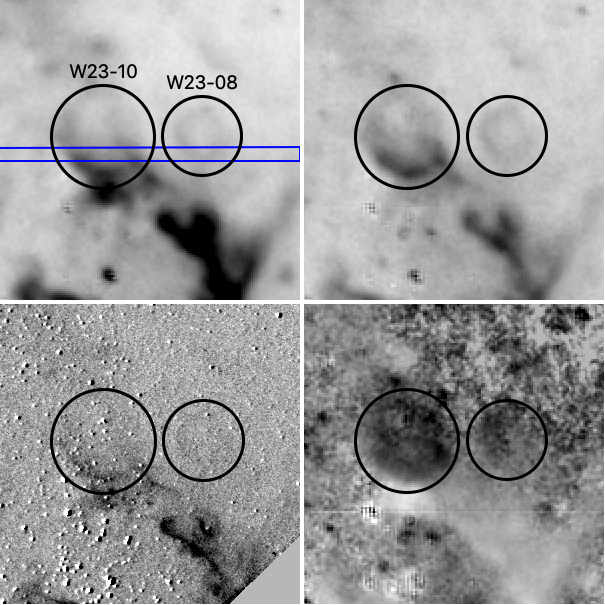}{0.38\textwidth}{(a)}
\fig{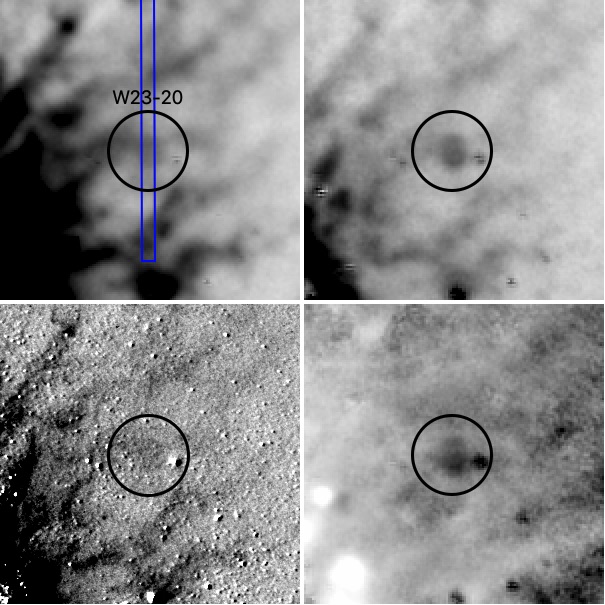}{0.38\textwidth}{(b)}
}
\vspace{-3mm}
\gridline{
\fig{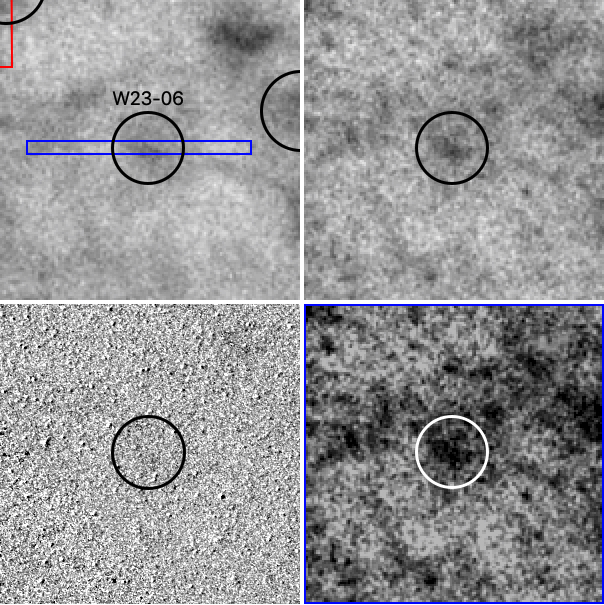}{0.38\textwidth}{(c)}
\fig{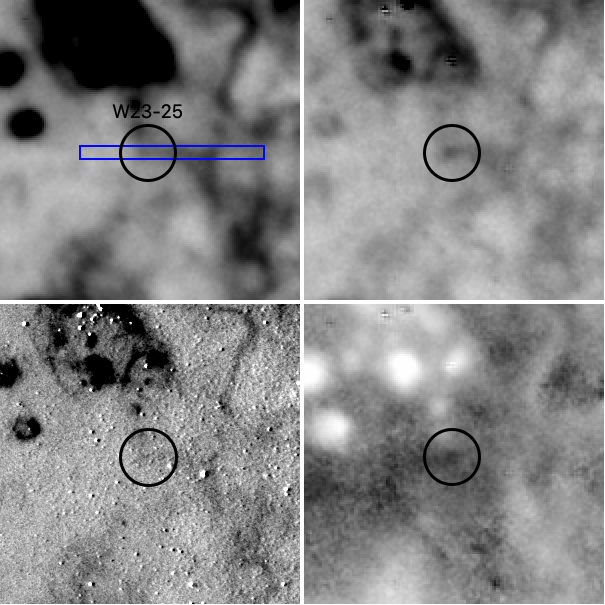}{0.38\textwidth}{(d)}
}
\vspace{-3mm}
\gridline{
\fig{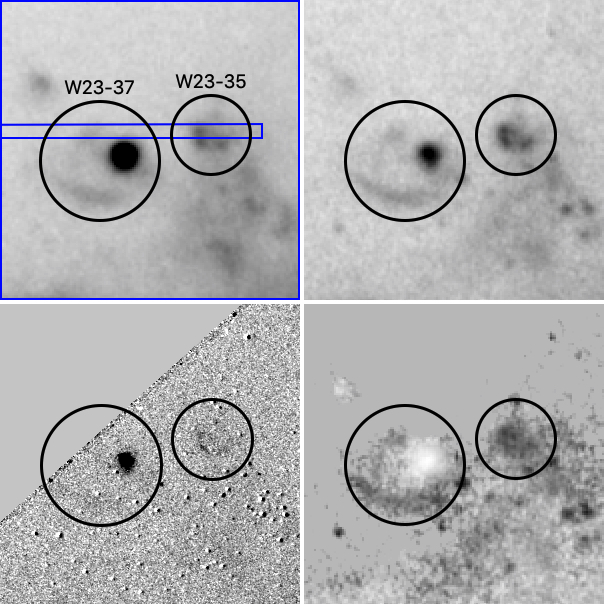}{0.38\textwidth}{(e)}
\fig{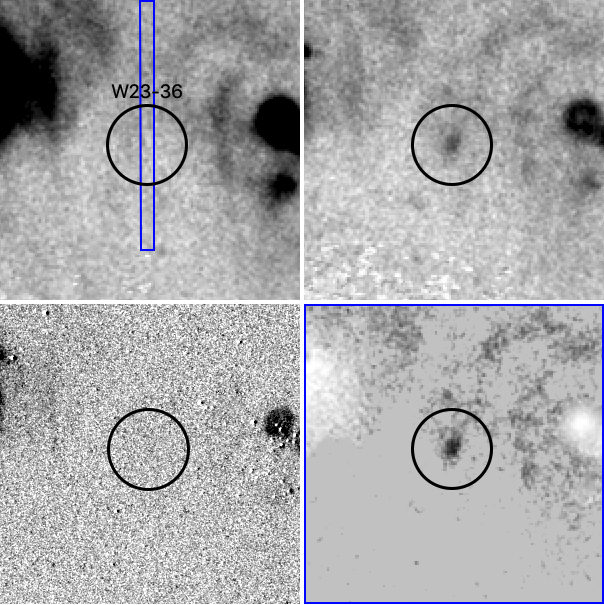}{0.38\textwidth}{(f)}
}
\vspace{-3mm}
\caption{(a) Images of a small field containing candidates W23-08 and W23-10 (both class A objects).  The panels are (clockwise from upper left) continuum-subtracted GMOS \ha; continuum-subtracted GMOS \sii; GMOS \sii:\ha\ ratio; {\em HST} ACS image in \ha\ (F658N), also continuum-subtracted.  Shown in blue on the \ha\ image is the slit used for the spectra of these objects, which crosses near the southern edge of both.  Darker in the ratio panel indicates higher ratio. The field is 15\arcsec\ square, oriented N-up, E-left.  (b) Same as panel a, except for  W23-22, class A; (c)  W23-06, a class B object; here the \sii:\ha\ ratio image shows that this entire field has diffuse emission with an elevated ratio; (d) W23-25, class B; (e)   W23-35 and W23-37, both class A, but note the contrast with the compact bright \hii\ region overlapping the larger and fainter W23-37; and (f) W23-36, class A, which shows up clearly in the \sii\ and ratio images, but does not stand out in \ha; note the contrast with the \hii\ region to the west. 
\label{fig:4panels}}
\end{figure} 

\subsection{Spectroscopy and Analysis}

Once these candidate objects had been identified, we prepared two GMOS masks
for follow-up multi-object spectroscopy, targeting a total of 30 candidates, plus the bright, O-rich SNR-1, two of the candidates 
from \citet{leonidaki13} that did not meet our selection criteria, and a handful of comparison
\hii\ regions.  We selected small faint \hii\ regions that were relatively isolated and comparable in surface brightness to the SNR candidates.  Some additional \hii\ region spectra were extracted from along several of the SNR apertures as well.  The \hii\ region positions are listed in Table 4. Having one mask with N-S slits and the other with E-W ones gave us additional flexibility in targeting objects in confused fields and in cases where we could target multiple objects on the same slit. These masks were used for the observations described above in Section 2.3.  Also included on Mask 1 was the bright SNR-1, which we will report on in a separate upcoming paper.

\begin{figure}
\gridline{
\fig{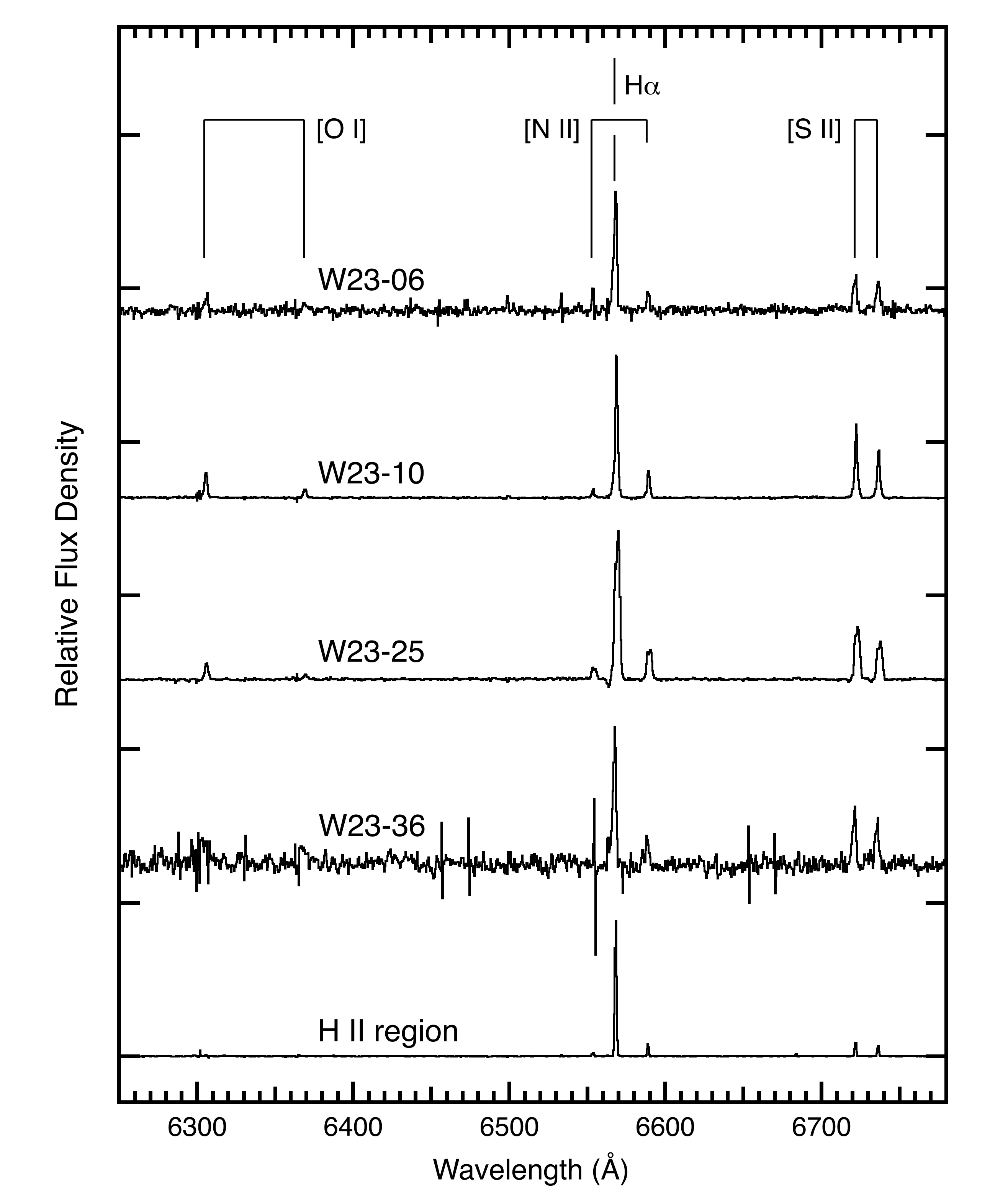}{0.684\textwidth}{(a)}
\fig{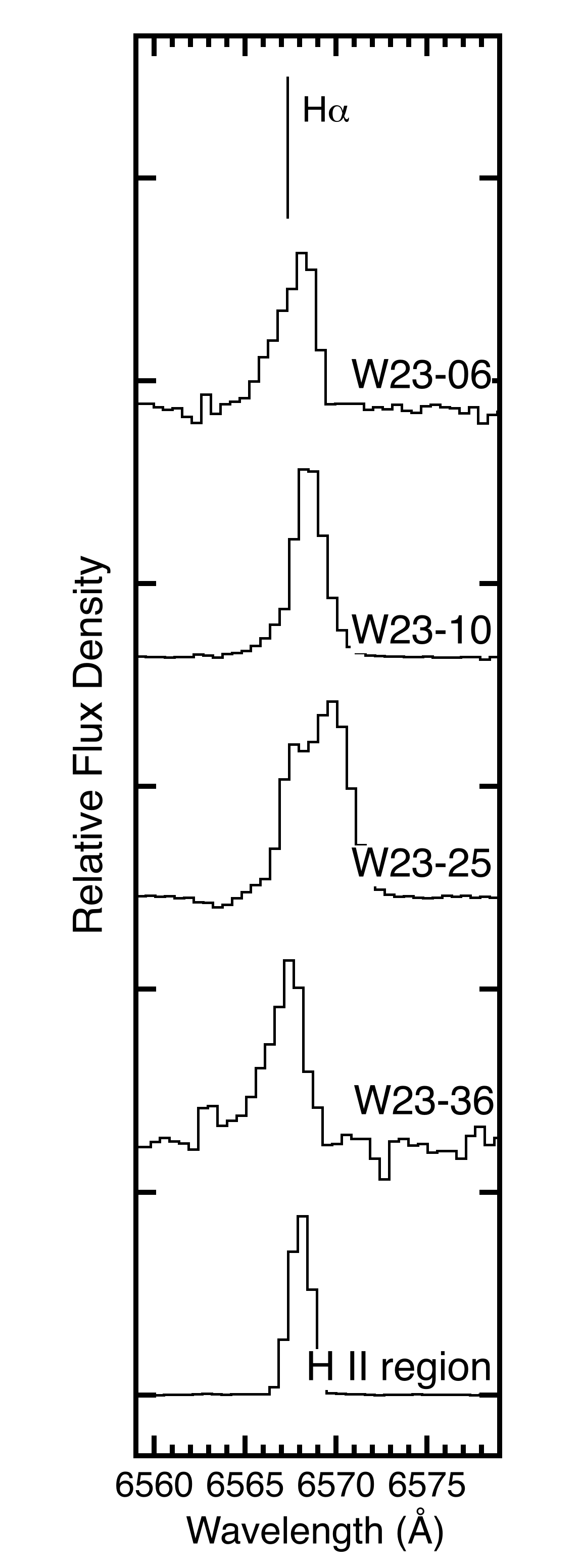}{0.30\textwidth}{(b)}
}
\caption{(a) Spectra of several SNRs in \gal, all ones shown in Fig.\ \ref{fig:4panels}, plus an \hii\ region for comparison. (b) An expanded version of a narrow range around
\ha\  for the same spectra, illustrating the velocity broadening for most SNRs, but not for the \hii\ region.  It is also evident that the assumption of Gaussian line shapes is an over-simplification; nevertheless, this gives a reasonable approximation of the line broadening and line fluxes.
\label{fig:spectra}}
\end{figure} 

In order to characterize the spectra in a relatively uniform manner, we  fit the most prominent lines in the spectra with simple Gaussians.  Prior to fitting, we inspected each of the spectra and set a flag that excluded obviously discrepant pixels from the fitting process.    
We  fit \OiL\ as a singlet, but, as was the case when we analyzed both the MUSE and GMOS spectra of M83 SNRs and \hii\ regions, we found that fitting \SiiLL\ as a doublet  and the \ha\,-\,\NiiLL\ complex as a triplet with fixed separations and a single FWHM (plus a constant background) produced more consistent results than fitting all of the lines individually, especially for the fainter spectra.  There were no cases where our visual inspections indicated any significant difference in the shapes of the \NiiLL\  compared to \ha, except for a few  that could be associated with a narrow component of \ha\ superposed on the broader profile of the SNR.  Several typical spectra are shown in Fig.~\ref{fig:spectra}.

The results of the spectral fits are presented in Tables \ref{snr_spectra}  and \ref{h2_spectra} for the SNRs and \hii\ regions in the sample, respectively.  The errors indicated in the tables should be treated with caution as they are statistical, and do not take into account systematic errors that might arise, for example, from imperfect background subtraction.  Such errors are difficult to quantify.

\section{Discussion \label{sec:discussion} }

\subsection{Overall characterization of the sample}

As we noted in Sec.\ \ref{sec:image_results}, the \sii:\ha\ ratio is generally high throughout \gal, so we expect this diagnostic to be of more limited utility in confirming SNRs than it has been in other galaxies, especially for objects whose \sii:\ha\ ratios are within $\sim$0.1 of the usual 0.4 criterion we use as a discriminant.  Our spectra confirm this prediction;  as shown in the histogram of \sii:\ha\ ratios in   Fig.\ \ref{fig:histograms} (a); while the bulk (28 out of 30) of the SNR candidates have \sii:\ha\ ratios that exceed the conventional 0.4 ratio, many (6 of 13) of our sample of faint \hii\ regions do as well. However, the objects that have \sii:\ha\ ratios in excess of $\sim$0.5 do seem to be significantly in excess of the surrounding diffuse gas.

\begin{figure}
\gridline{
\fig{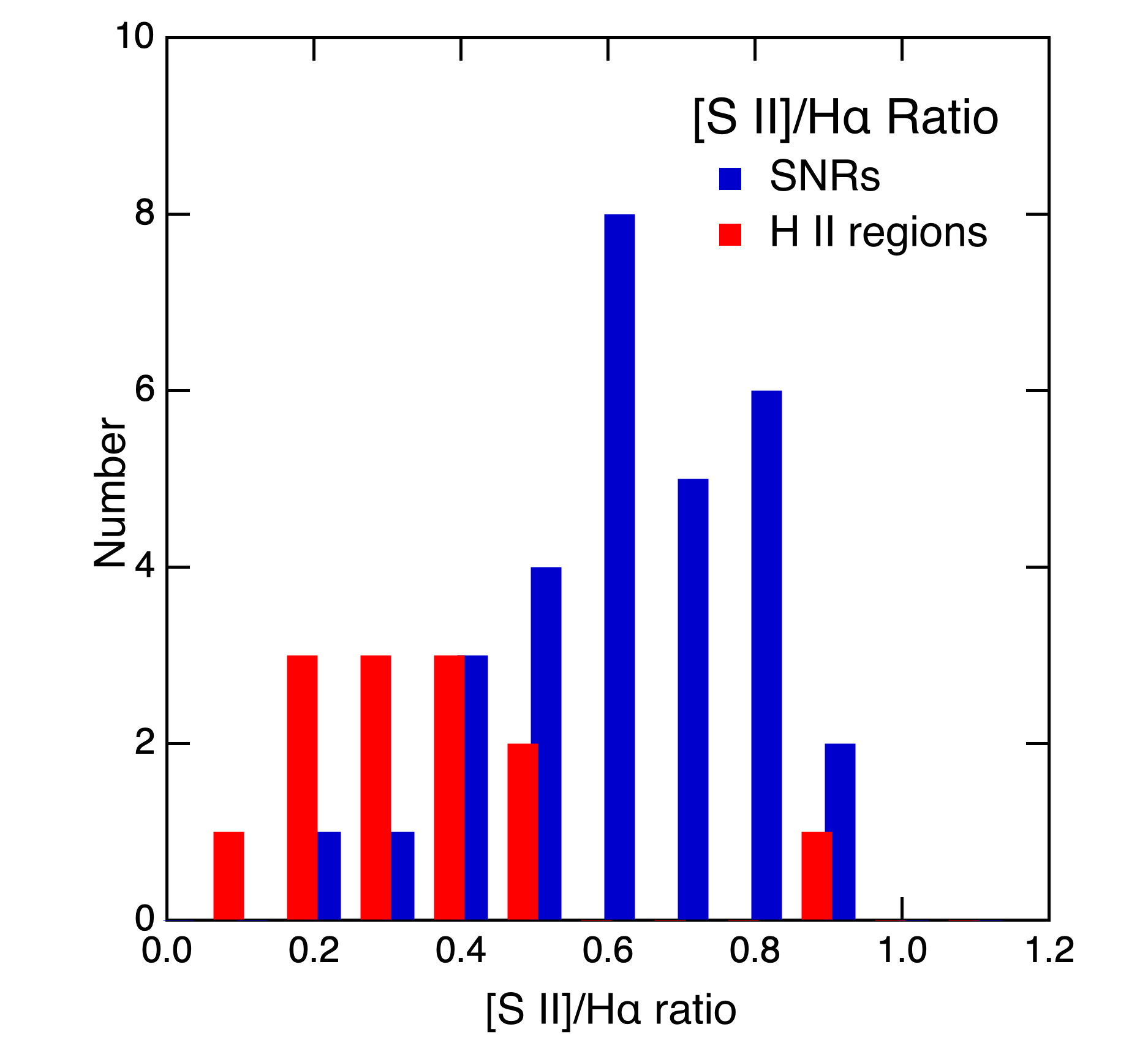}{0.34\textwidth}{(a)}
\fig{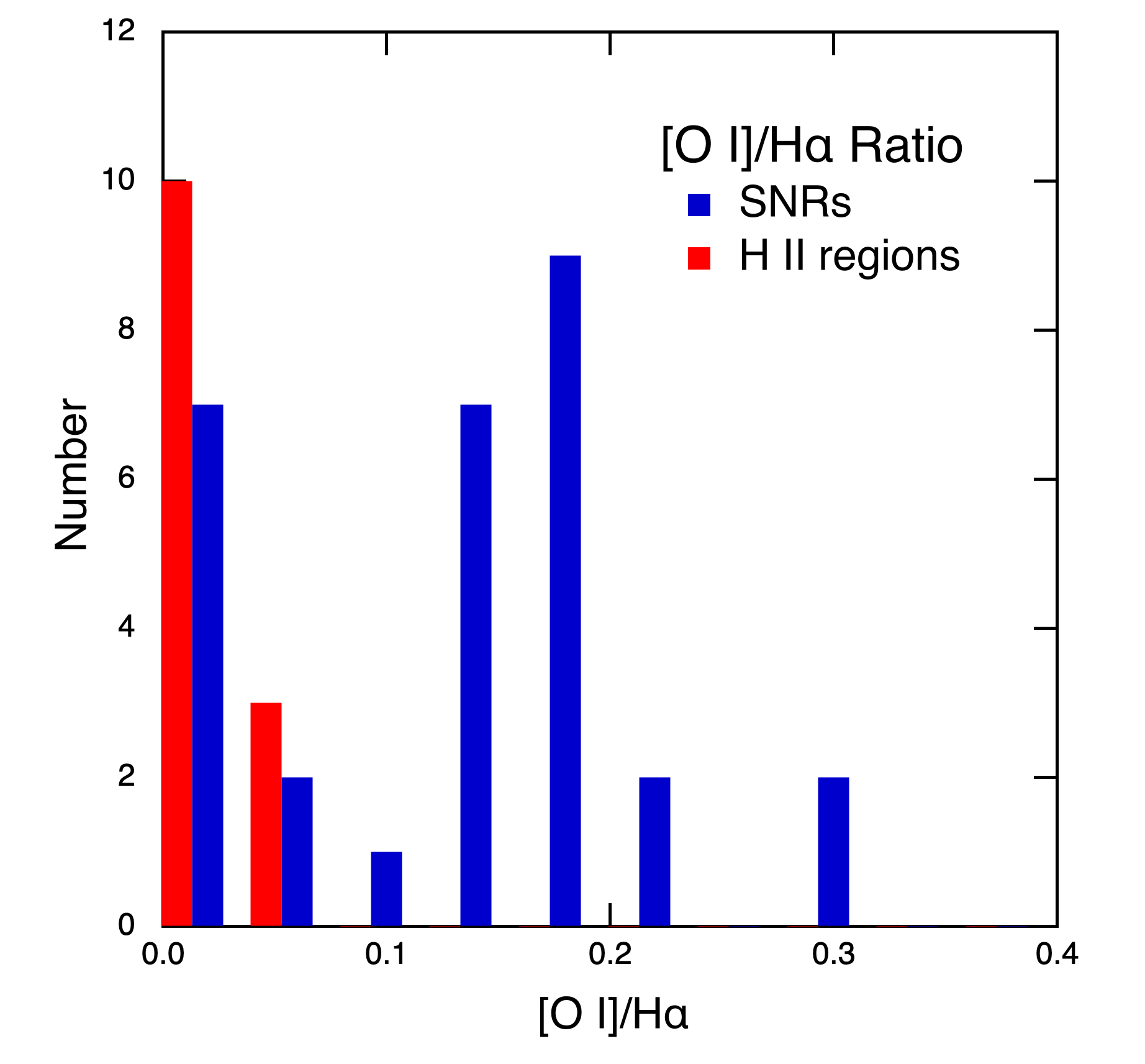}{0.34\textwidth}{(b)}
\fig{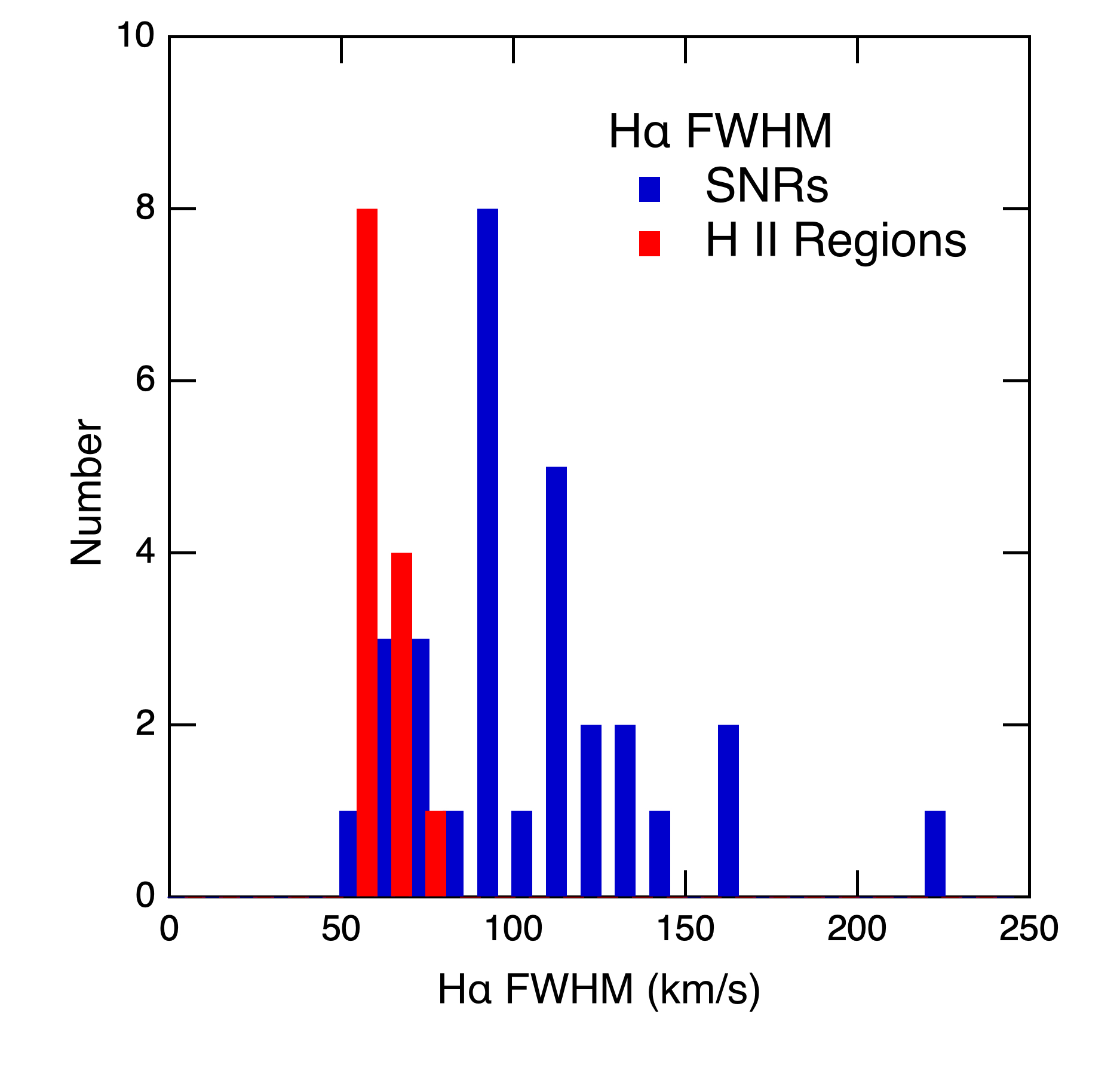}{0.34\textwidth}{(c)}
}
\caption{(a) Histogram of the \sii:\ha\ ratio, as measured spectroscopically for objects identified through imaging as SNR candidates or \hii\ regions. 
This line ratio seems  to be a less useful diagnostic for SNRs in \gal\ than in most other galaxies.  (b) Similar histogram as at left, but for the \oi:\ha\ ratio, which gives  better separation between the two classes of objects.  (c) Similar histogram as those at left, but for the Gaussian width of the \ha\ line.  The line width seems to give the best separation between the two types of objects.
\label{fig:histograms}}
\end{figure}

A much cleaner separation is obtained using the FWHM of the emission lines; 22 of the 30 SNR candidates with spectra have emission line widths of $\ge$90 $\kms$, while none of the \hii\ regions has a velocity width this high.  This does not imply that some of the SNR candidates with narrower lines are {\em not} SNRs, but does imply that nearly all of the objects with large velocity widths are SNRs. We note that the discriminator of 90 $\kms$ is larger than one might expect for an instrumental resolution of 65$\kms$ and typical \hii\ region velocity dispersion of order 10 - 20 $\kms$, but the statistical quality of the data and the limitations of simple Gaussian fitting lead us to choose the higher velocity threshold.  

Are there other line ratios that would be of use in discriminating between SNRs and photoionized gas?  \cite{kopsacheili20} have argued that the \oi:\ha\ ratio is an additional strong indicator of SNR shocks, and that any nebula with \oi:\ha\ $>0.01$ is a probable SNR\@.  In M83, \cite{winkler23} found that most SNRs had  \oi:\ha\ ratios of 0.03 or greater (limited in part by sensitivity), arguing that at least for M83,  confident detection of \oi\ should be considered when identifying SNRs optically.  Of course, \oi\ can be affected by night sky subtraction, and \gal\ is  redshifted by only $\sim$200 $\VEL$ ($\sim$4.3 \AA).  As shown in Fig.\  \ref{fig:histograms}(b), there is a clear separation in this ratio between most of our SNR candidates and the faint \hii\ regions for which we have data.  Rather conservatively, we adopt a cutoff of 0.10 for \gal: all SNRs have \oi:\ha\ $>0.10$, while no \hii\ regions do.


Fig.\ \ref{fig:fwhm_comp} shows the correlation between the line widths for the  \sii\ lines and that for \ha.  We primarily use the line width for \ha\ because it is the strongest line, but as this figure shows, using the \sii\ width would give essentially the same results.

\begin{figure}[h!]
\plotone{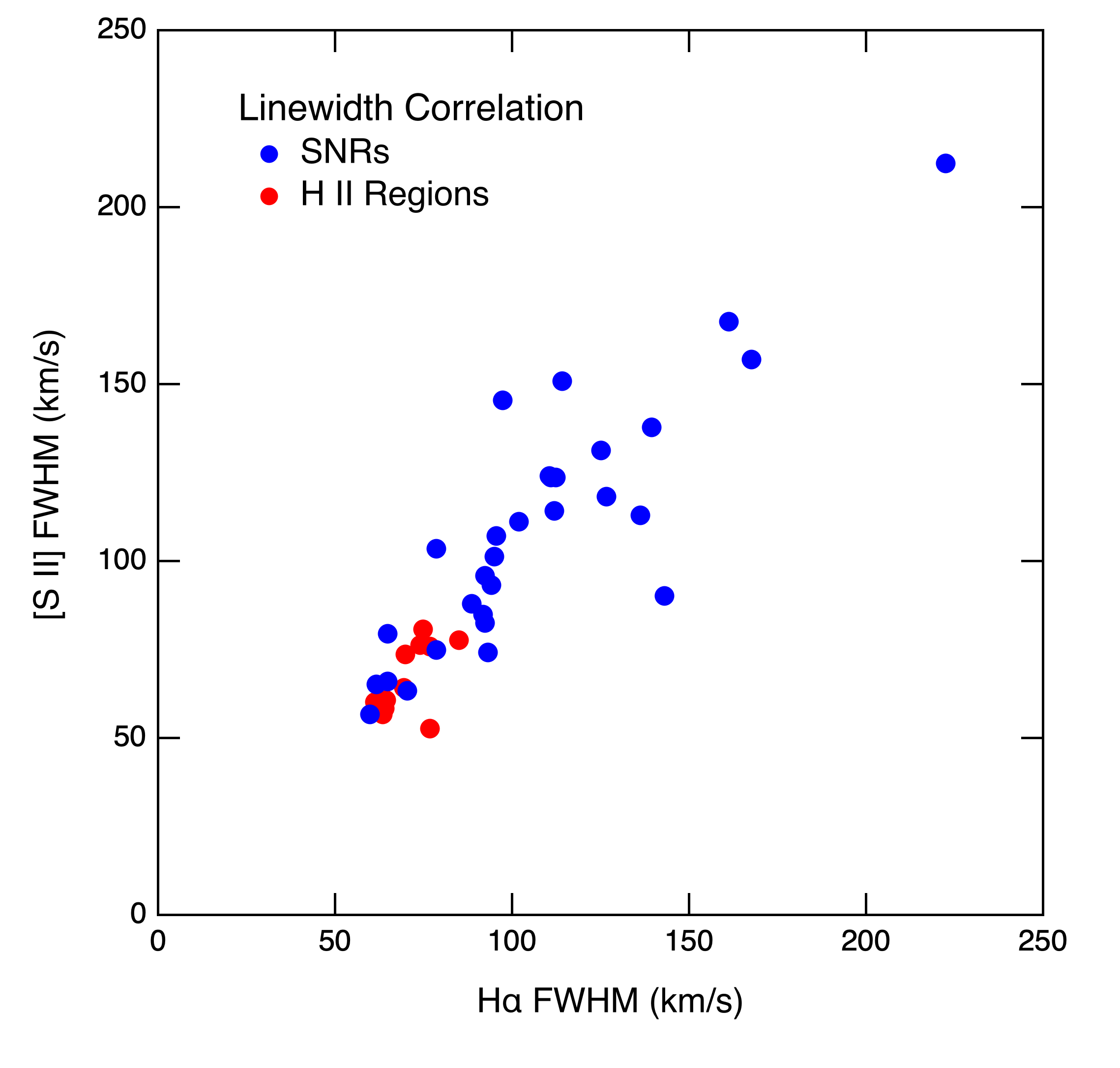}
\caption{Correlation between the line width for the \sii\ lines and that for \ha.  There is little difference between the measured line widths for these lines.
\label{fig:fwhm_comp}}
\end{figure}

Comparison among our three diagnostics are shown graphically in Fig.\ \ref{fig:fwhm_vs_ratios} and Fig.\ \ref{fig:compare}(a), and  a summary of the the numbers of spectroscopically-observed SNRs and \hii\ regions that pass various tests is presented in Table \ref{stats}.  Of the 30 SNR candidates for which we have spectra, 25 have \sii:\ha\ ratios  higher than a conservative value of 0.5, and 15 pass all three of our tests:  ratios that exceed our thresholds for both \sii:\ha\ and \oi:\ha\  and have profiles with FWHM of greater than 90 $\VEL$. These 15 objects are almost certainly SNRs. 
The \oi:\ha\ and \sii:\ha\ ratios are fairly well correlated, as shown in Fig.~\ref{fig:compare}(a),  indicating that candidates that meet {\em either} ratio criterion {\em and} are velocity-broadened are most likely to be SNRs.  
As shown in panels (b) and (c) of Fig.~\ref{fig:compare}, there is little if any correlation of the \nii:\ha\ ratio with either \sii:\ha\ or \oi:\ha.

\begin{figure}[h!]
\plottwo{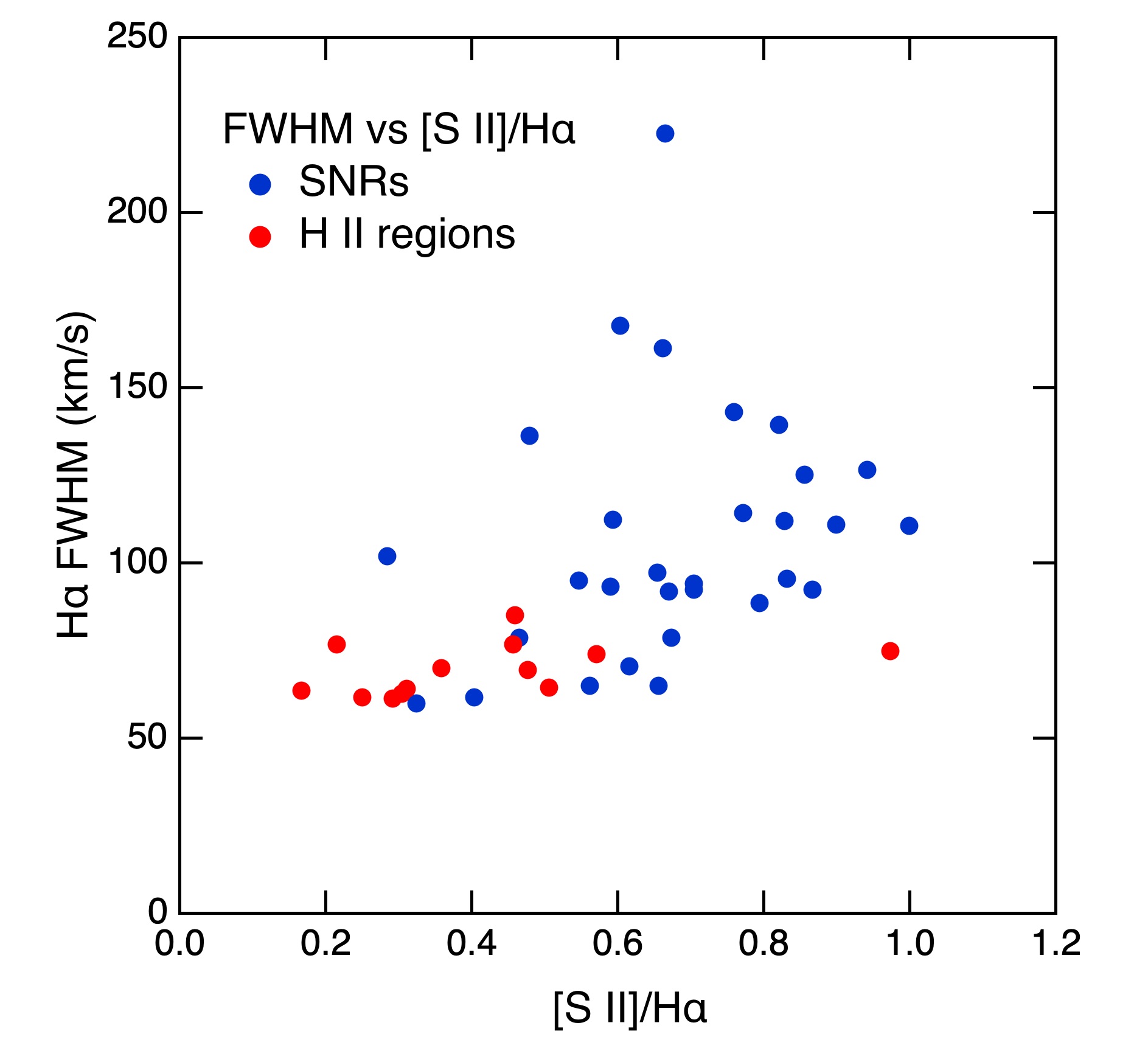}{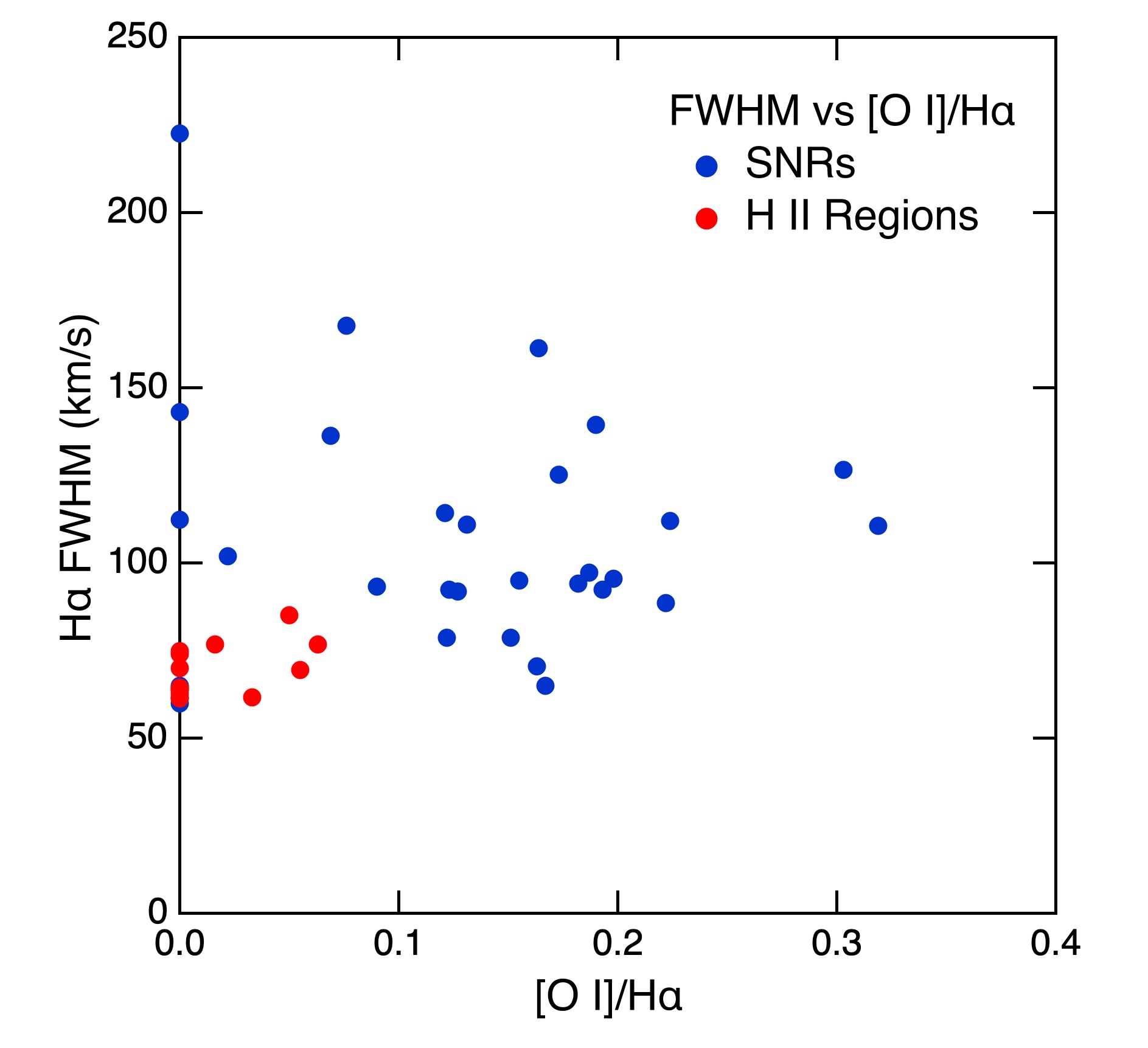}
\caption{Comparison between different diagnostics for distinguishing between SNRs and \hii\ regions.  ({\em left}) Line width vs.\ the \sii:\ha\ ratio.   ({\em right}) Line width vs.\ the \oi:\ha\ ratio.
\label{fig:fwhm_vs_ratios}}
\end{figure} 

\begin{figure}
\gridline{\fig{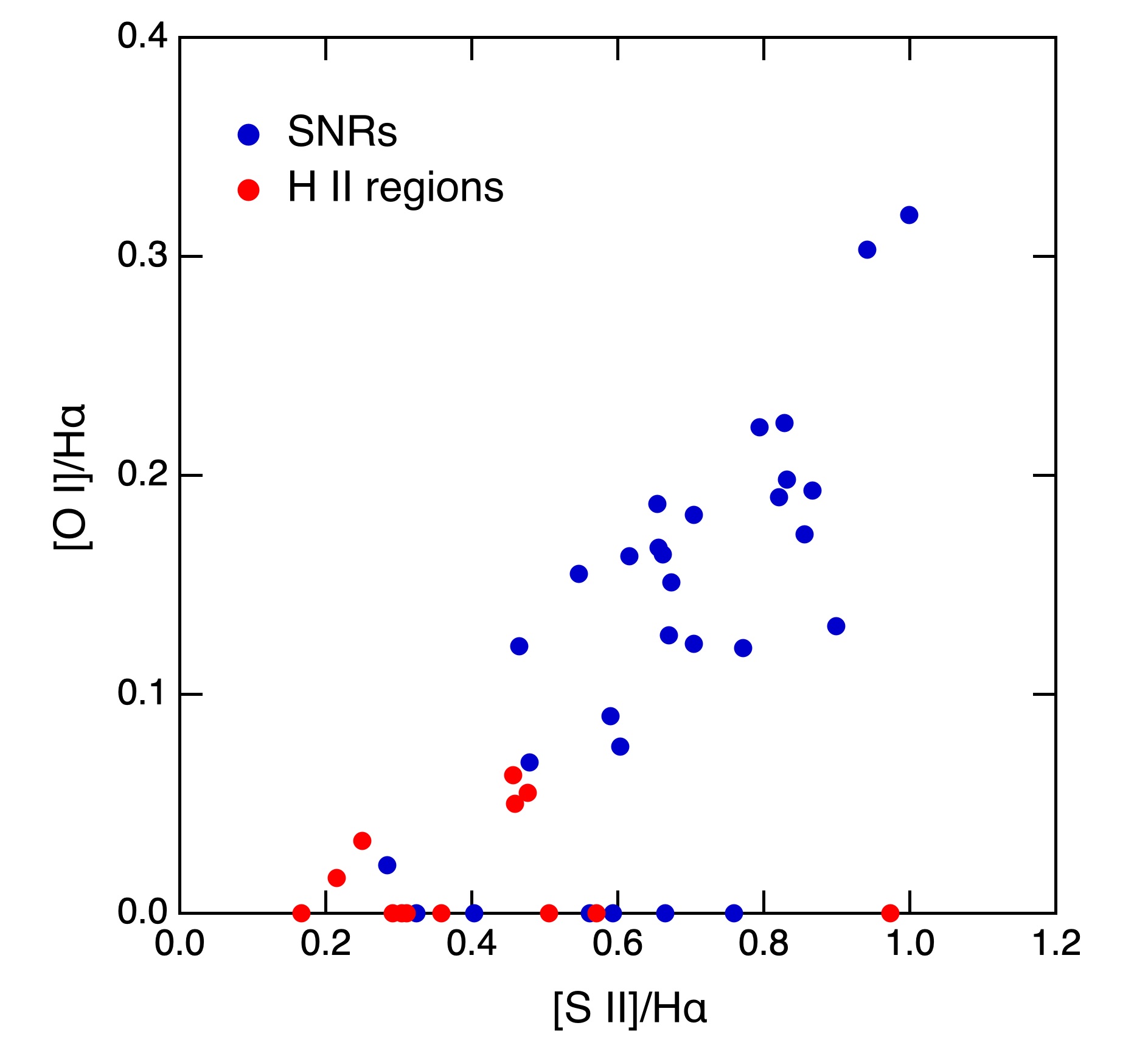}{0.34\textwidth}{(a)}
          \fig{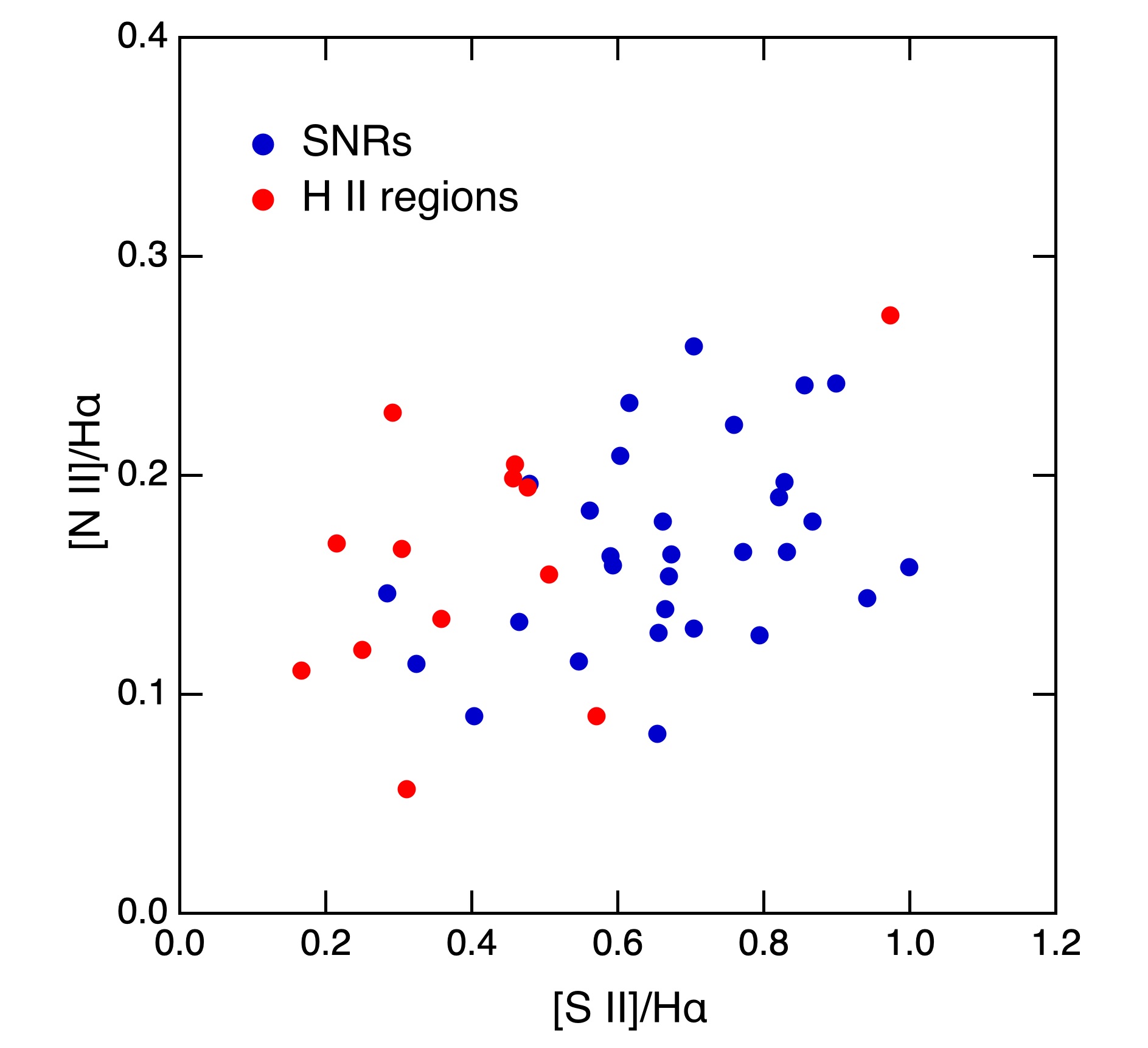}{0.34\textwidth}{(b)}
          \fig{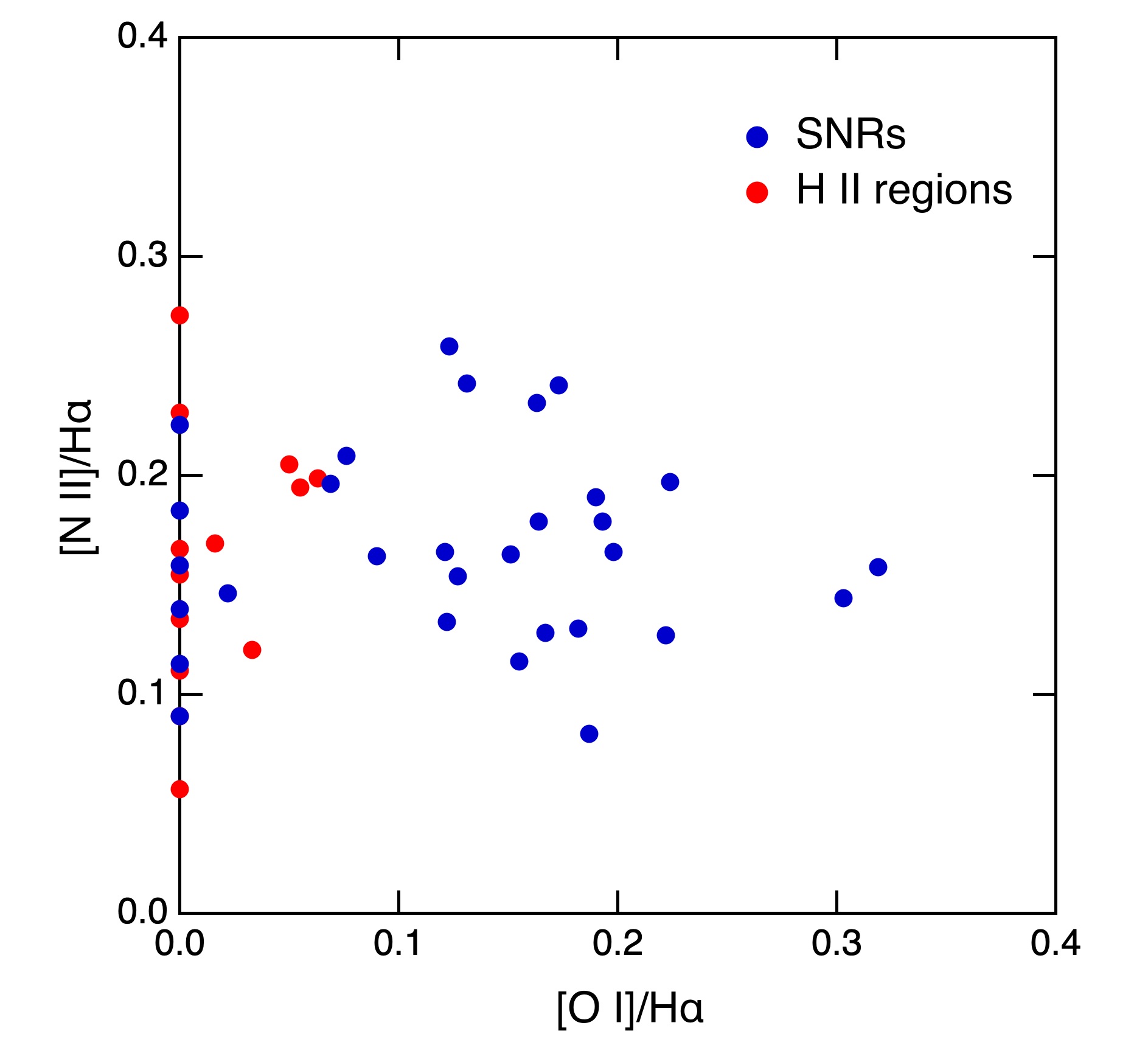}{0.34\textwidth}{(c)}}
\caption{(a) Comparison between the \oi\,$\lambda\,6300$ and the \SiiLL\ line strengths (both relative to \ha) in \gal.  (b) A similar comparison but for the \NiiLL\ and \SiiLL\ lines.  
(c) A similar comparison but for the \NiiLL\ and  \oi\,$\lambda\,6300$  lines.  
Only for the \oi\ and \sii\ lines (a) does there appear to be any correlation. \label{fig:compare} 
 } 
\end{figure}

Within the SNR sample there is  little difference between the morphology class A and class B objects, in terms of the percentage of objects with spectra that satisfy the various tests, with the exception that the morphology class B objects tend not to have high \oi:\ha\ ratios.  Since \oi\ is produced as post-shock gas cools to temperatures lower than required for \sii, one could speculate that objects with well-defined morphologies have a higher percentage of emission from complete shocks than those with less well-defined morphologies. 

As expected,  in part because of how we have tailored the cuts, {\em positive} tests for the \hii\ regions differ substantially from those for SNRs;   None of the \hii\ regions shows \oi:\ha\  greater than 0.1 nor a FWHM of greater than 90 $\VEL$, and only three show a \sii:\ha\ ratio greater than 0.5. This latter point indicates that, even with the elevated \sii:\ha\ criterion used here, we cannot be entirely confident in identifying some objects.

\subsection{Our sample vs.\ that of \citet{leonidaki13} }
Our sample of SNRs and candidates in \gal\ comprises 49 objects, while Leonidaki's has 71, yet the two samples have merely four objects in common.  This is unexpected in view of the fact that both our group and theirs selected candidates using the same basic method: narrow-band imaging in \ha\ and \sii, and selecting objects with high \sii:\ha\ ratio relative to the surrounding background.

There are, nevertheless, important differences in our techniques.  Notably, the GMOS images  from the 8.1\,m Gemini-N telescope that we used for our final selection have significantly higher sensitivity and better seeing (0\farcs 6 - 0\farcs 7) than those used by \citet{leonidaki13}, which were taken from the 1.3\,m Skinakas telescope in seeing from 1\farcs 3 to 2\farcs 5\@.  Furthermore, \citet{leonidaki13} selected their candidates using an automated technique, SExtractor, while we selected ours by careful visual examination of the various imagery noted in Sec.\  2.\footnote{While automated selection is probably valuable in many cases where the \sii:\ha\ criterion is more effective, we expect the visual inspection technique described here to be superior in the case of \gal.}  We further required our candidates to be nebulae with individual integrity that  stood out as relatively isolated objects, brighter than the surrounding background. This was strictly true for our class A candidates, but by definition the class B ones were more marginal in this regard.

Any selection based on the \sii:\ha\ ratio in \gal\ requires further confirmation, since this ratio is unusually high in the diffuse emission that pervades much of the galaxy, and especially in the more central regions where many of the \citet{leonidaki13} objects were identified.  Our selection of nebulae that clearly stood out as individual objects, and our use of additional spectroscopic diagnostics,  all help to mitigate confusion caused by this generally elevated \sii:\ha\ ratio.   In Fig.\ \ref{fig:3panels} we show images of several representative \citet{leonidaki13} objects: the four that we also identified (panels e - h; see Table 3) and four others not included in our list, selected because \citet{leonidaki13} observed particularly high photometric \sii:\ha\ ratios for all: LBZ-1 (\sii:\ha\ = 0.93), LBZ-12 (0.77), LBZ-13 (0.63), and LBZ-14 (0.79).  
The absence of identifiable objects with elevated ratio at the positions of the non-selected objects (panels a - d) should be clear.

\begin{figure}
\gridline{
\fig{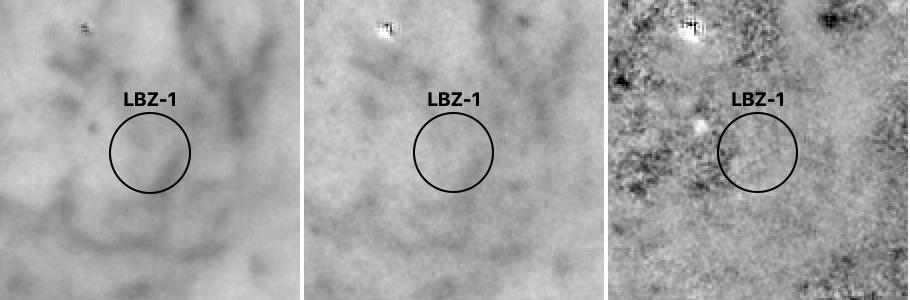}{0.49\textwidth}{(a)}
\fig{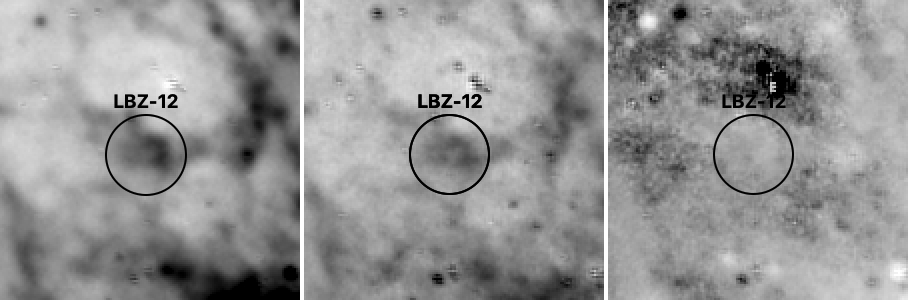}{0.49\textwidth}{(b)}
}
\vspace{-1mm}
\gridline{
\fig{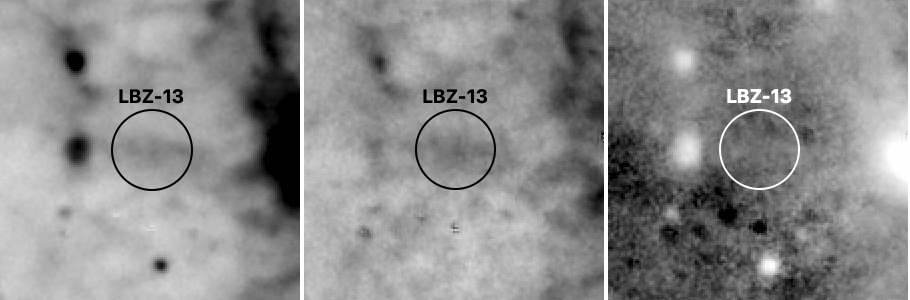}{0.49\textwidth}{(c)}
\fig{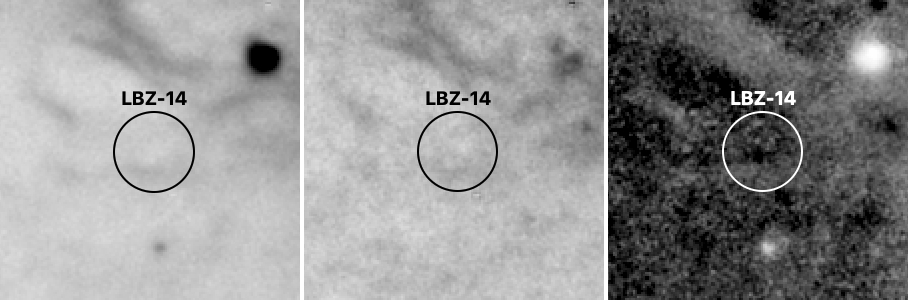}{0.49\textwidth}{(d)}
}
\vspace{-1mm}
\gridline{
\fig{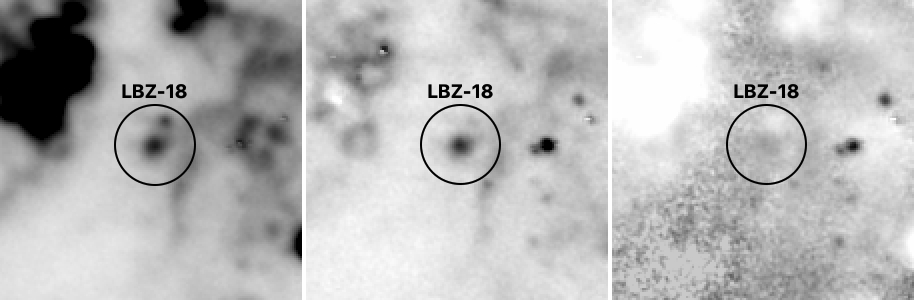}{0.49\textwidth}{(e)}
\fig{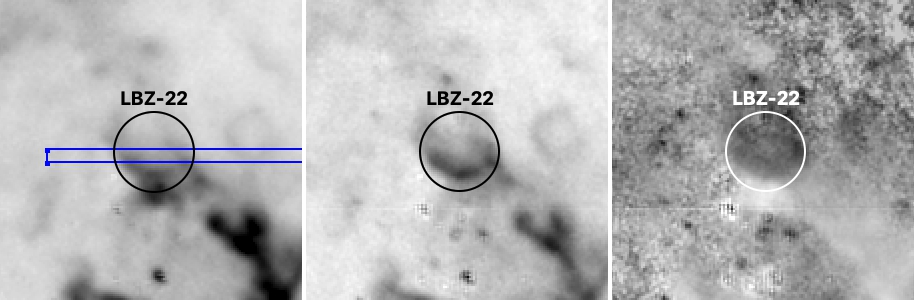}{0.49\textwidth}{(f)}
}
\vspace{-1mm}
\gridline{
\fig{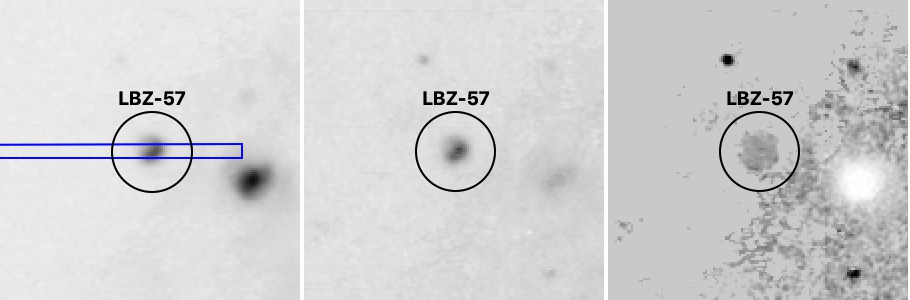}{0.49\textwidth}{(9)}
\fig{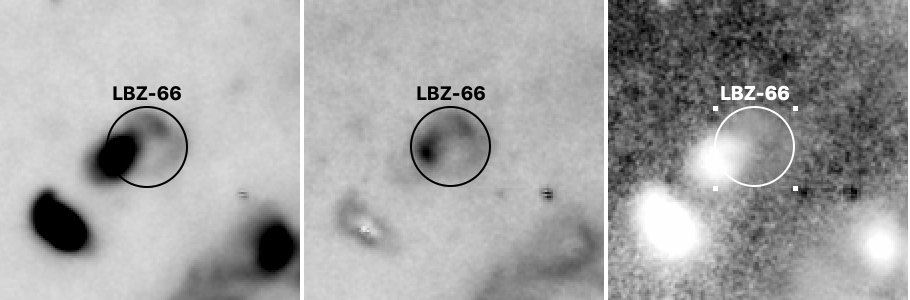}{0.49\textwidth}{(h)}
}
\vspace{-1mm}

\caption{(a) Images of several small fields, each containing one of the LBZ candidates.  The sub-panels are (left to right) continuum-subtracted GMOS \ha; continuum-subtracted GMOS \sii; GMOS \sii:\ha\ ratio. The grey scale is different for different objects, but the {\em relative} scaling for \ha\ and \sii\ is the same for all.  For all the \sii:\ha\ ratio sub-panels, the scaling is identical and darker means higher ratio.    Each small field is 15\arcsec\ square, oriented N up, E left, and each small circle is 4\arcsec\  in diameter.  Panels a-d all show LBZ candidates that did {\em not} meet our selection criteria.  We do not see objects identifiable as possible,SNRs in our images, which are deeper and have better seeing than those of \citet{leonidaki13}.  The objects in panels e-h are all the ones that we selected as well:  LBZ-18 = W23-46, LBZ-22 = W23-10, LBZ-57 = W23-45, LBZ66 = W23-34.   For objects from which we obtained spectra, our slit positions are shown in blue on the \ha\ images.   
 \label{fig:3panels}}
\end{figure}

For spectroscopic confirmation of candidates, we used slits of width 0\farcs 6, which served to effectively isolate our objects from the surrounding emission, in addition to facilitating our desired velocity resolution.  By contrast, the spectra reported by \citet{leonidaki13} used  a slit of width either 6.3\arcsec,
or 2.5\arcsec\ for the objects observed at KPNO\@.
Because of the high ratio in the diffuse gas and the large apertures used, it is hardly surprising that they measured high \sii:\ha\ ratios in virtually all the objects they observed. 

We obtained spectra of four of the \citet{leonidaki13} objects---two of which were on our candidate list, and two of which were not.  Of these, LBZ22 and LBZ57 pass all three of our tests, and are almost certainly SNRs.   We did not obtain spectra of the other two objects in common between our list and theirs, LBZ18 and LBZ66, so these remain good candidates.  However, spectra of 
LBZ49 and LBZ67 pass none of our criteria, and these should no longer be regarded as good candidate SNRs.

\subsection{Diffuse background emission in \gal}

\begin{figure}
\includegraphics[width=1.0\textwidth]{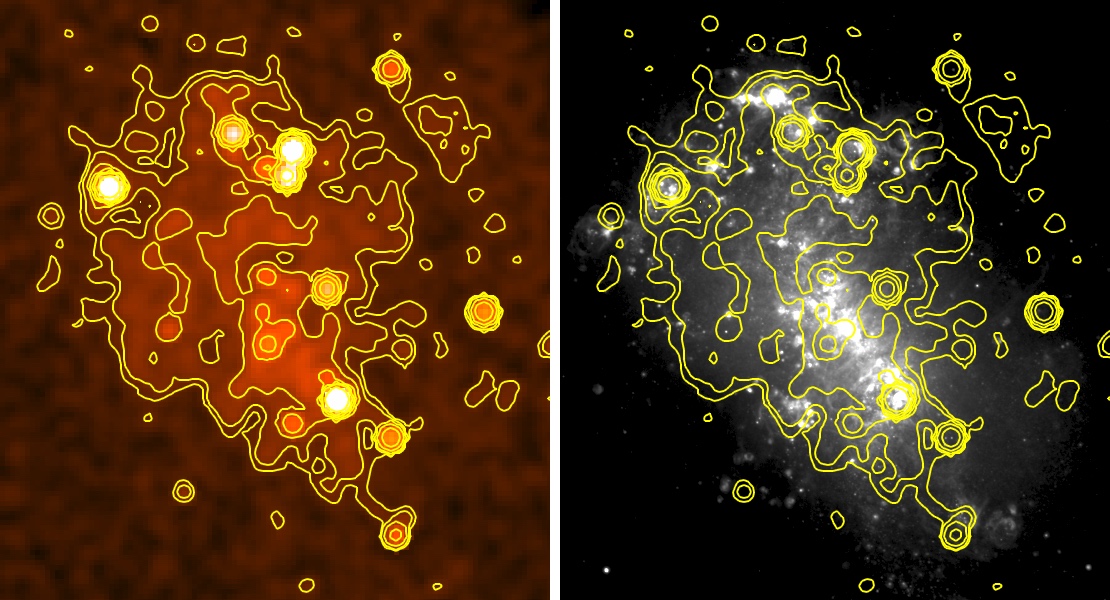}
\caption{
({\em left}) {\em Chandra}  0.5 - 4.0 keV X-ray image of \gal, along with contours for the same image.   ({\em right})  \ha\ image from GMOS, with the same X-ray contours superimposed. The diffuse \ha-emitting gas is roughly coincident with the much hotter X-ray plasma \citep{summers03}.
\label{fig:xray_pair}}
\end{figure}

As discussed by \cite{summers03},  {\em Chandra} X-ray images of \gal\  show not only a number of compact X-ray sources, but also diffuse emission whose morphology resembles the optically-emitting  gas. The pervasiveness of \sii\ emission in \gal\ of course means that there is low-ionization gas throughout much of the galaxy.  It is interesting to compare the optical emission with the soft X-ray emission  that also pervades much of the galaxy (Fig.~\ref{fig:xray_pair}; see also Fig.~\ref{fig:s2_by_ha}).
Both the X-rays and low-ionization optically emitting gas are probably related to the intense star formation that is taking place in \gal\ \citep{annibali08, karczewski13, calzetti18, sacchi18}, quite possibly triggered by a merger with a dwarf galaxy \citep{martinez12}.  The population of very hot, blue stars is surely producing intense winds in their vicinity.  Since the mass of \gal\ is low, the winds are not gravitationally confined, and they drive significant outflows from \gal\ \citep{hong13, bomans14, mcquinn19}.  The UV emission from the hot stars and their winds can in turn excite diffuse gas to X-ray temperatures, and also produce shocks that lead to the low-ionization optical emission. The X-ray gas is understood to be the hot component of an outflow driven by multiple SN and stellar winds arising from young massive stars in the Galaxy. \gal\ was one of the first Irr galaxies where such outflows were identified \citep{summers03}.

Dwarf galaxies with starbursts are favorable locations for such outflows to exist because of their low gravitational potential.  The low metallicity of dwarf starbursts is at least partly due to the fact that they cannot retain the metals that are produced by SN explosions \citep{dalcanton07}.   These outflows are mass-loaded; that is, they contain some amount of hot gas but also very significant amounts of molecular gas. \sii:\ha\ ratios can be high in such outflows as a result of shocks that develop within the outflowing gas \citep{sharp10}.   
The fact that \gal\ contains such an outflow and that these outflows are brighter than the diffuse interstellar gas (DIG) seen in other galaxies is the underlying reason why identifying SNRs in \gal\ has been so difficult.

\subsection{The SNR Sample in \gal\ Compared with Other Galaxies}

There are now numerous galaxies in which samples of several dozen to several hundred likely SNRs have been identified \citep[][and references in the Introduction]{vucetic15}.  The list of observed galaxies is heavily weighted toward spirals.
Not surprisingly given their relative masses, the spirals with well-characterized SNR samples at distances comparable to that of \gal\ or closer  all harbor significantly larger SNR populations.  Fig.~\ref{fig:5gals} shows a comparison of the \nii:\ha\ ratios for \gal\ and four well-studied spirals, all plotted as a function of \sii:\ha.  All are similar in terms of their \sii:\ha\ ratios (except that those for \gal\ are somewhat higher, as we have discussed).  But the \nii:\ha\ ratios differ systematically by as much as an order of magnitude.  Such dramatic differences reflect the differences in metallicity of the galaxies.  As was the case in the other galaxies, most, but not all, of the SNR candidates have \SiiLL\ line ratios near the low density limit of about 1.4\@.

\begin{figure}
\plotone{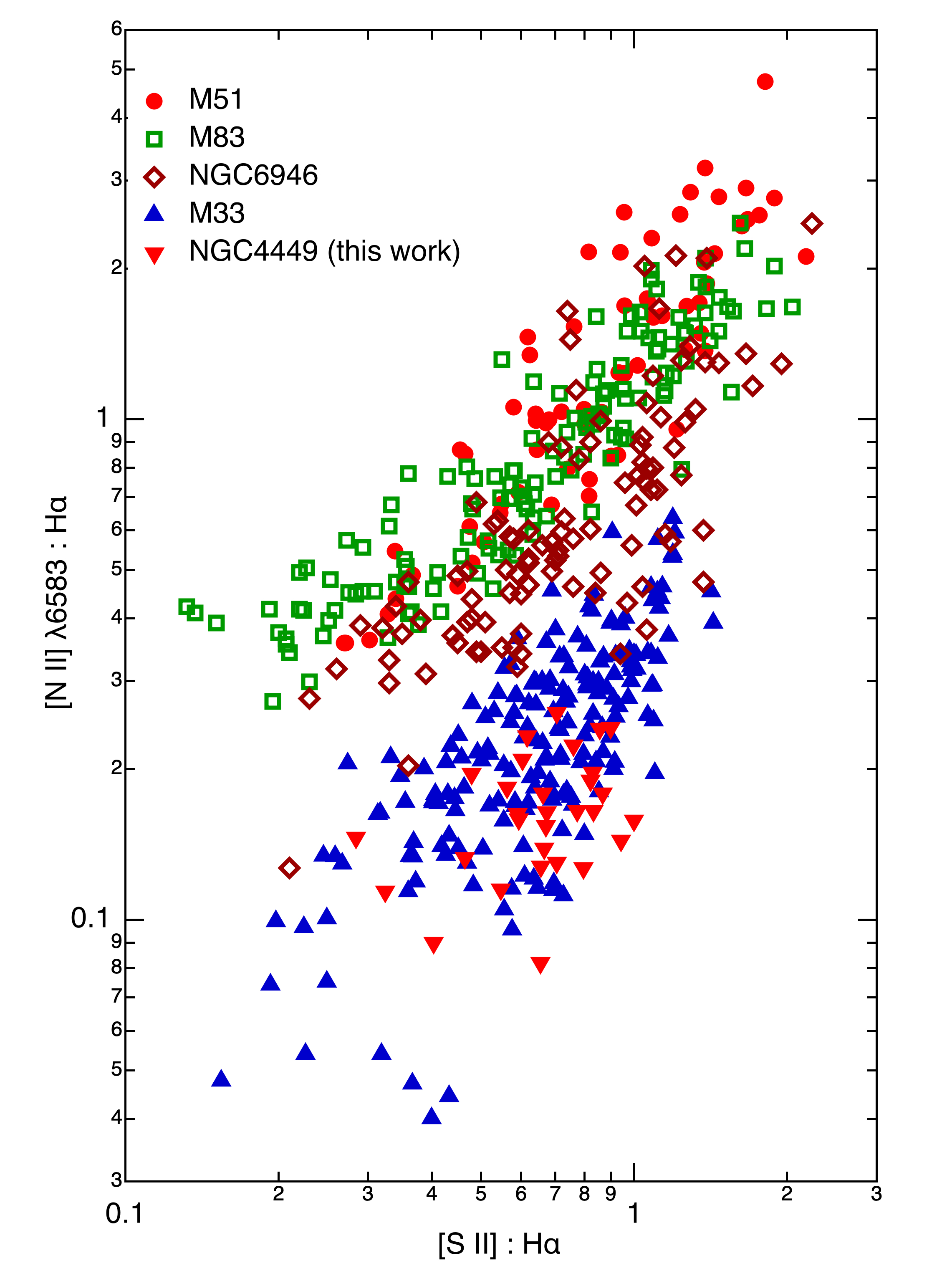}
\caption{Log-log plot comparing the \nii:\ha\ ratio and the \sii:\ha\ ratio for SNRs in \gal\ with the results for four spiral galaxies.  \nii\ is weaker in \gal\ than in the other galaxies except for M33, reflecting the low metallicity of both M33 and \gal.  References: M33: \citet{long18};  M51: \citet{winkler21}; M83: \citet{winkler17}; NGC\,6946: \citet{long19}.
\label{fig:5gals}}
\end{figure}

The SNRs and candidates in \gal\ range from a minimum diameter of 11.5 pc to a maximum of 147 pc, with a median of 36 pc.  Comparing this with other spiral galaxies, we find that the comparable statistics for M51 and M83 are both considerably smaller (medians of 24 pc and 20.6 pc, respectively), while for M33 they are larger (median 51 pc).  It is reasonable that larger objects are seen in M33; it is far closer and better resolved than \gal, with less diffuse background emission.  Thus we can detect older, lower surface-brightness (and thus larger) SNRs in M33 than in \gal.  The M33 sample also includes numerous partial shells, which have been extrapolated to estimate the diameter.  Such objects, if present, could have been easily missed in our \gal\ search.  
Most comparable to \gal\ is the Large Magellanic Cloud, where the (probably nearly complete) sample of 59 confirmed SNRs has a median diameter of 36 pc \citep{ou18}, almost identical to \gal.  Given the proximity of the LMC and the quality of the available multiwavelength data sets, it is not surprising that the number of confirmed SNRs is significantly higher than for \gal. 

The SNR sample in \gal, like those of the other galaxies, is dominated by older SNRs, close to or already entering the radiative phase.  This is expected, both because optical searches using the \sii:\ha\ ratio select SNRs with emission from radiative shocks (with recombination timescales less than the age of the SNR), and the total \ha\ (and \sii) luminosity of such shocks from a SNR typically increases with object diameter, simply because the surface area of the SNR increases with time.

\subsection{How Complete is the \gal\ Sample?}

\citet{li11} have developed a model for estimating the supernova rate for a galaxy as a function of its B-band and K-band luminosity.  Applying their formalism to \gal\ yields an estimate of 0.35 SNe per century, essentially all of which should arise from core-collapse explosions.  If we take a typical age of 20,000 years for a SNR to fade significantly in brightness as it begins to merge into the ISM, we find that \gal\ might have been expected to have of order 70 visible SNRs.  As argued above, we have detected 15 to 30 of these (16 to 31, if we also include SNR-1).  

Thus, it is reasonable to suppose that we have detected a sizable fraction of the potentially visible SNRs in \gal, but certainly not all of them.  Similar arguments lead to a similar conclusion for the larger spiral galaxies with the best-studied samples of SNRs, viz. M33 and M83, that is that we are only seeing a sub-sample of the total population of SNRs that may be present. This could be due to several effects.  First of all, we know there are some remnants of Type Ia SNe that are seen optically as faint hydrogen Balmer-line emitters rather than bright radiative shocks.  Any such objects would certainly have been missed.  Secondly, in rapidly star forming galaxies, SNe that occur within the cavities blown by stellar winds and previous SNe are not expected to produce bright optical remnants as the shocks have little to run into.  Thirdly, we note that there are a number of LMC SNRs that appear as `blisters' or partial shells on the edge of bright \hii\ regions.  Such objects could easily be hidden or missed in the complex emission seen in \gal.

\subsection{Searching for SNRs in Starburst Galaxies: the Path Forward}

The bright diffuse emission in galaxies similar to \gal\ and the fact that this diffuse emission can have elevated line ratios for both \sii:\ha\ (and probably \oi:\ha\ as well) makes identifying SNRs from low-resolution optical spectroscopy problematic, except in those cases where high spatial resolution imagery reveals the distinct morphology expected for an isolated SNR, and where a local background can be robustly subtracted.
Three paths forward seem warranted:

(a) SNRs are demonstrated to have velocity widths that are broad compared to those of \hii\ regions (cf. \citet{points19}. Hence, catalogs constructed of resolved optical objects with broad lines are likely to be SNRs.   Integral field spectroscopy, with instruments such as MUSE on the VLT or  SITELLE on the CFHT, and with sufficient spectral resolution to measure velocity broadening $\lesssim 30 \kms$,  could provide a powerful technique for this application. \citep[The recent study by][provides an illustration of what might be achieved.]{congiu23}
Even the 65 $\kms$ resolution obtained in our study leaves an ambiguity for some SNRs whose velocity dispersions may  be high compared with \hii\ regions,  but that are still not velocity-resolved.  Of course, obtaining even higher spectral resolution for faint extended nebular sources can be challenging.

(b) SNRs also differ from \hii\ regions and diffuse gas seen in outflows in terms of their radio spectral indices.  SNRs are steep-spectrum objects, whereas radio sources powered by free-free emission have flat spectra.  Most Galactic SNRs were discovered on the basis of this distinction. However, extending such observations to  galaxies beyond the Local Group requires  both spatial resolution and sensitivity at multiple frequencies that have been difficult to achieve \citep[cf.][]{russell20}. Nevertheless, instruments such as the JVLA or the forthcoming Square Kilometre Array offer great potential.

(c) A third possibility, probably even more challenging, is higher sensitivity X-ray telescopes with {\em Chandra}-like spatial resolution and good spectral resolution. The detection of individual objects with spectra that are unambiguously thermal would not only positively identify SNRs, but also serve to reveal additional information about them.  Unfortunately, X-ray telescopes with all of these properties seem not to be on the immediate horizon.

\section{Summary and Conclusions}

We have identified 49 nebulae in \gal\  as SNR candidates, based on narrow-band imaging in \sii\ and \ha\ lines, that shows identifiable objects with elevated \sii:\ha\ ratios compared with their surroundings.  We have obtained spectra of 30 of the candidates, confirming the high \sii:\ha\ ratios in most of the objects.  All remain viable SNR candidates, and most show additional features that support their identification as SNRs.  There are 15 objects that  meet all three criteria that we propose for SNRs in \gal:  \sii:\ha\ ratios greater than 0.5, \oi:\ha\ ratios  greater 0.1, and \ha\  line widths greater than 90 $\VEL$.  We contend that these 15, at minimum, are almost certainly SNRs.

Our list of SNRs and SNR candidates is almost disjoint from a set of 71 objects proposed  by \citet{leonidaki13}.  Despite our doing a targeted search for nebulae at the positions they had identified, we are able to spectroscopically confirm only two of their objects as almost certain SNRs, and two more (for which we do not have spectra) as strong SNR candidates.  For the majority of their objects, and especially those in the complex inner regions of the galaxy, we do not see objects identifiable as potential SNRs in our images, which are deeper and have significantly better seeing than the earlier data.

The SNRs and SNR candidates that we have found are not special compared to the SNRs that are found in other galaxies, aside from the fact that the spectra reflect the low abundances in \gal.  Based on their size distribution, most are older SNRs, likely  entering the radiative phase, when SNRs are brightest at optical wavelengths.

Assuming  \gal\ is typical of other irregular galaxies with the same general morphology, about 70 SNe should have exploded in the last 20,000 years, almost all of them core-collapse.  While the age of the SNRs in our sample is uncertain, and the detectability of SNRs depends on the environment into which they are  expanding, this suggests that we have found a significant fraction of the remnants from SNe that have exploded during that period, but clearly this list is incomplete. 

Further progress in the identification and study of SNRs in galaxies beyond the Local Group is possible via integral field spectroscopy with instruments such as MUSE, especially if the kinematic resolution is sufficient to separate shocked gas from photoionized regions.  Deep surveys in radio or possibly X-ray could, in principle, also help identify SNRs using different criteria.  Such work is needed in any event to thoroughly characterize both individual SNRs and their population as a whole.  

\vspace{0.4in}

Based on observations obtained at the international Gemini Observatory, a program of NSF's NOIRLab, which is managed by the Association of Universities for Research in Astronomy (AURA) under a cooperative agreement with the National Science Foundation on behalf of the Gemini Observatory partnership: the National Science Foundation (United States), National Research Council (Canada), Agencia Nacional de Investigaci\'{o}n y Desarrollo (Chile), Ministerio de Ciencia, Tecnolog\'{i}a e Innovaci\'{o}n (Argentina), Minist\'{e}rio da Ci\^{e}ncia, Tecnologia, Inova\c{c}\~{o}es e Comunica\c{c}\~{o}es (Brazil), and Korea Astronomy and Space Science Institute (Republic of Korea).
We thank the staff at Gemini-North for their excellent support throughout several semesters during which weather, equipment problems, and the pandemic disrupted the observations.

PFW acknowledges financial support from the National Science Foundation through grants AST-0908566 and AST-1714281\@. WPB acknowledges support from the Dean of the Krieger School of Arts and Sciences and the Center for Astrophysical Sciences at JHU during this work.

\facilities {Gemini:North (GMOS), WIYN  3.5m, HST (ACS)}

\software {SAOimage ds9, IRAF}

\bibliographystyle{aasjournal}

\bibliography{bibmaster}


\begin{deluxetable}{llccrr}[hb!]
\tablewidth{0pt}
\tablecaption{Imaging Observations of NGC\,4449}

\tablehead{
 \colhead{} &  \colhead{} &\multicolumn{3}{c}{Filter} & \colhead {}\\ 
\cline{3-5}  
\colhead{Telescope} &
 \colhead{Date} &\colhead{Designation} &
 \colhead{$\rm \lambda_{c}$(\AA)} &
\colhead{$\Delta \lambda$(\AA)\tablenotemark{a}} &
\colhead {Exposure (s)\tablenotemark{b,c}} 
}

\startdata
WIYN 3.5m & 2011 Jun 26-28 & \oiii &  5010 &60\phn\phn & $3\times800$ \phn\phn  \\
& & Green Continuum  & 5127& 100\phn\phn&$3\times500$ \phn\phn \\
& & H$\alpha$  & 6563 & 27\phn\phn    & $3\times800$ \phn\phn \\
& & \sii\tablenotemark{d} & 6723 & 63\phn\phn & $3\times800$ \phn\phn  \\
& & Red Continuum  & 6840& 93\phn\phn &$3\times600$ \phn\phn \\\\
Gemini-N 8.1m & 2021 Apr 6 & H$\alpha$  & 6576 & 70\phn\phn    & $8\times300$ \phn\phn \\
& & \sii\ & 6719 & 52\phn\phn & $8\times600$ \phn\phn  \\
& & r & 6311& 1357\phn\phn &$8\times60$ \phn\phn \\
\enddata

\tablenotetext{a}{Full width at half maximum.}
\tablenotetext{b}{Number of exposures $\times $ individual exposure time.}
\tablenotetext{c}{Half of the Gemini (GMOS) exposures were taken at P.A. 90\degr, and half at P.A. 180\degr\ to give flexibility in MOS mask design.}
\tablenotetext{d}{WIYN Observatory filter W037; other filters used at WIYN were PFW custom ones.}

\label{imaging_obsns}
\end{deluxetable}

\begin{deluxetable}{cccccc}
\tablewidth{0pt}
\tablecaption{Gemini-N/GMOS Spectroscopy Observations of \gal \label{tab:journal}}
\tablehead{
\colhead{Mode} &
\colhead{Grating} &
\multicolumn{2}{c}{Center Position} &
\colhead{Date (UT)}  &
\colhead{Total Exposure (s)\tablenotemark{a} }\\
\colhead{}  & \colhead{} & \multicolumn{2}{c}{R.A.\phn\,(J2000)\, Decl.} & \colhead{} & \colhead{}
}
\tablewidth{0pt}
\startdata
MOS Mask 1, N-S slits & B1200 & 12:28:08.52 \phn & 44:05:32.00   & 2022 Mar 28, Jun 5  & 3 CWLs\,$\times \,3 \times 1000$ \\[5pt]
MOS Mask 2, E-W slits & B1200 &12:28:09.54\phn\phn & 44:05:33.94    & 2022 Apr 1, Jun 5   & 3 CWLs\,$\times\, 3 \times 1000$ \\
\enddata
\tablenotetext{a}{Number of different Central Wavelength (CWL) Settings $\times$ number of exposures at each CWL $\times$ individual exposure time (s).}
\end{deluxetable}

\begin{deluxetable}{lrrrccrrrl}
\tabletypesize{\scriptsize}
\decimals
\tablecaption{Properties of Supernova Remnants and Candidates in NGC\,4449$^a$}
\vspace{-2 mm}
\tablehead{
\colhead{Source$^b$} &
\colhead{R.A.} &
\colhead{Decl.} &
\colhead{Diameter} &
\colhead{Morphology} &
\colhead{Spectrum} &
\colhead{\sii:H$\alpha >$ 0.5} &
\colhead{\oi:H$\alpha >$0.1} &
\colhead{FWHM$>$90 km s$^{-1}$} &
\colhead{Other name(s)$^c$}   \\ [-2ex]
\colhead{} &
\colhead{(J2000)} &
\colhead{(J2000)} &
\colhead{(pc)} &
\colhead{Class} &
\colhead{} &
\colhead{} &
\colhead{} &
\colhead{} &
\colhead{}
}
\startdata
W23-01 & 12:27:54.58 & 44:05:13.7 & 147 & A & Yes & No & No & No & -- \\ 
{\bf W23-02} & 12:27:54.79 & 44:05:30.3 & 98 & A & Yes & Yes & Yes & Yes & -- \\ 
W23-03 & 12:28:04.65 & 44:05:08.7 & 37 & B & Yes & Yes & No & Yes & -- \\ 
W23-04 & 12:28:04.83 & 44:04:23.2 & 31 & A & -- & -- & -- & -- & -- \\ 
W23-05 & 12:28:04.87 & 44:04:56.6 & 54 & B & -- & -- & -- & -- & -- \\ 
W23-06 & 12:28:05.58 & 44:04:54.8 & 44 & B & Yes & Yes & No & Yes & -- \\ 
W23-07 & 12:28:06.24 & 44:05:03.0 & 44 & B & Yes & Yes & No & Yes & -- \\ 
W23-08 & 12:28:06.55 & 44:04:30.9 & 52 & A & Yes & Yes & Yes & No & -- \\ 
W23-09 & 12:28:06.60 & 44:03:41.2 & 38 & B & -- & -- & -- & -- & -- \\ 
{\bf W23-10} & 12:28:07.00 & 44:04:30.7 & 79 & A & Yes & Yes & Yes & Yes & LBZ22 \\ 
W23-11 & 12:28:07.23 & 44:04:04.4 & 48 & B & Yes & Yes & No & No & -- \\ 
W23-12 & 12:28:07.43 & 44:05:32.4 & 64 & A & -- & -- & -- & -- & -- \\ 
{\bf W23-13} & 12:28:07.65 & 44:04:09.9 & 71 & A & Yes & Yes & Yes & Yes & -- \\ 
{\bf W23-14} & 12:28:07.81 & 44:03:59.7 & 60 & A & Yes & Yes & Yes & Yes & -- \\ 
W23-15 & 12:28:07.84 & 44:04:25.3 & 33 & A & -- & -- & -- & -- & -- \\ 
W23-16 & 12:28:08.75 & 44:03:57.3 & 39 & B & Yes & No & No & No & -- \\ 
W23-17 & 12:28:09.15 & 44:05:25.2 & 22 & A & -- & -- & -- & -- & -- \\ 
W23-18 & 12:28:09.68 & 44:05:19.3 & 21 & A & Yes & No & No & Yes & CW09-07 \\ 
W23-19 & 12:28:09.74 & 44:04:06.3 & 62 & A & -- & -- & -- & -- & -- \\ 
{\bf W23-20} & 12:28:10.32 & 44:05:42.6 & 33 & A & Yes & Yes & Yes & Yes & -- \\ 
W23-21 & 12:28:10.66 & 44:07:16.5 & 57 & A & -- & -- & -- & -- & -- \\ 
W23-22 & 12:28:11.09 & 44:05:37.1 & 23 & A & Yes & No & No & Yes & nuclear source \\ 
W23-23 & 12:28:11.48 & 44:05:36.4 & 12 & A & Yes & No & Yes & No & S03-18; CW09-15 \\ 
W23-24 & 12:28:12.12 & 44:07:34.1 & 117 & A & -- & -- & -- & -- & -- \\ 
W23-25 & 12:28:12.37 & 44:05:18.6 & 20 & B & Yes & Yes & No & Yes & -- \\ 
W23-26 & 12:28:12.84 & 44:07:06.8 & 24 & B & -- & -- & -- & -- & -- \\ 
W23-27 & 12:28:13.00 & 44:05:16.8 & 31 & A & -- & -- & -- & -- & -- \\ 
{\bf W23-28} & 12:28:13.10 & 44:05:37.7 & 29 & A & Yes & Yes & Yes & Yes & CW09-19 \\ 
W23-29 & 12:28:13.30 & 44:05:14.5 & 60 & A & -- & -- & -- & -- & -- \\ 
W23-30 & 12:28:13.38 & 44:07:18.5 & 25 & B & -- & -- & -- & -- & -- \\ 
W23-31 & 12:28:13.44 & 44:07:31.3 & 65 & A & Yes & Yes & Yes & No & -- \\ 
W23-32 & 12:28:13.69 & 44:07:17.9 & 19 & B & -- & -- & -- & -- & -- \\ 
W23-33 & 12:28:13.69 & 44:06:10.6 & 24 & B & -- & -- & -- & -- & -- \\ 
W23-34 & 12:28:14.45 & 44:05:17.1 & 66 & A & -- & -- & -- & -- & LBZ66 \\ 
{\bf W23-35} & 12:28:14.70 & 44:07:30.1 & 33 & A & Yes & Yes & Yes & Yes & -- \\ 
{\bf W23-36} & 12:28:14.95 & 44:04:32.3 & 32 & A & Yes & Yes & Yes & Yes & S03-24 \\ 
W23-37 & 12:28:15.24 & 44:07:28.5 & 76 & A & Yes & Yes & Yes & No & -- \\ 
{\bf W23-38} & 12:28:15.28 & 44:07:13.8 & 36 & A & Yes & Yes & Yes & Yes & -- \\ 
{\bf W23-39} & 12:28:15.29 & 44:06:18.1 & 14 & A & Yes & Yes & Yes & Yes & -- \\ 
W23-40 & 12:28:15.59 & 44:06:28.4 & 17 & A & Yes & Yes & No & Yes & -- \\ 
{\bf W23-41} & 12:28:17.61 & 44:06:36.6 & 25 & B & Yes & Yes & Yes & Yes & -- \\ 
{\bf W23-42} & 12:28:18.24 & 44:06:38.7 & 32 & B & Yes & Yes & Yes & Yes & -- \\ 
W23-43 & 12:28:18.33 & 44:06:06.8 & 36 & B & -- & -- & -- & -- & -- \\ 
W23-44 & 12:28:18.43 & 44:06:20.2 & 26 & A & Yes & Yes & Yes & No & -- \\ 
{\bf W23-45} & 12:28:19.25 & 44:06:55.4 & 33 & A & Yes & Yes & Yes & Yes & LBZ57; CW09-26 \\ 
W23-46 & 12:28:19.59 & 44:06:13.6 & 19 & A & -- & -- & -- & -- & LBZ18 \\ 
{\bf W23-47} & 12:28:21.07 & 44:06:07.8 & 44 & A & Yes & Yes & Yes & Yes & -- \\ 
{\bf W23-48} & 12:28:22.53 & 44:05:34.2 & 52 & A & Yes & Yes & Yes & Yes & -- \\ 
W23-49 & 12:28:22.79 & 44:06:40.0 & 45 & A & -- & -- & -- & -- & -- \\ 
\hline
LBZ-49$^d$ & 12:28:14.20 & 44:05:10.1 & -- & -- & Yes & No & No & No & -- \\ 
LBZ-67$^d$ & 12:28:14.90 & 44:04:45.9 & -- & -- & Yes & No & No & No & -- \\ 
\enddata
\vspace{-0.5mm}
\tablenotetext{a}{Not included in the table is the bright, O-rich SNR-1, located at RA = 12:28:10.93, Dec = 44:06:48.5.   It is a bright source at all wavelengths: S03-15 and CW09-12.}
\vspace{-1.5mm}
\tablenotetext{b}{Objects in bold pass all three spectroscopic tests, and hence are almost certainly SNRs.}
\vspace{-1.5mm}
\tablenotetext{c}{Previously identified objects within 2\arcsec\ of a SNR candidate.  LBZ objects refer to optical SNR candidates identified by \cite{leonidaki13}; S03 objects 
refer to X-ray sources (not necessarily SNR candidates) identified by \cite{summers03}; and  CW09 objects refer to radio SNR candidates identified by \cite{chomiuk09a}. }
\vspace{-1.5mm}
\tablenotetext{d}{ \cite{leonidaki13} candidates for which we obtained spectra.}
\label{snr_master}
\end{deluxetable}

\begin{deluxetable}{lrr}
\tabletypesize{\scriptsize}
\decimals
\tablecaption{Comparison \hii\ Regions in NGC~4449}
\tablehead{
\colhead{Source} &
\colhead{RA} &
\colhead{Dec}
\\
\colhead{} &
\colhead{(J2000)} &
\colhead{(J2000)}
}
\startdata
H~II-01 & 12:27:54.58 & 44:05:25.1 \\ 
H~II-02 & 12:27:55.60 & 44:05:26.6 \\ 
H~II-03 & 12:27:56.41 & 44:04:53.3 \\ 
H~II-04 & 12:27:56.72 & 44:05:25.7 \\ 
H~II-05 & 12:27:59.21 & 44:05:25.7 \\ 
H~II-06 & 12:28:01.98 & 44:06:11.2 \\ 
H~II-07 & 12:28:03.84 & 44:05:20.4 \\ 
H~II-08 & 12:28:08.70 & 44:03:10.6 \\ 
H~II-09 & 12:28:08.78 & 44:03:28.2 \\ 
H~II-10 & 12:28:10.94 & 44:06:51.9 \\ 
H~II-11 & 12:28:11.07 & 44:07:04.1 \\ 
H~II-12 & 12:28:16.65 & 44:06:13.7 \\ 
H~II-13 & 12:28:21.31 & 44:05:56.9 \\ 
\enddata
\label{h2_master}
\end{deluxetable}

\begin{deluxetable}{rrrrrrrrr}
\tabletypesize{\scriptsize}
\decimals
\tablecaption{NGC4449 candidate SNR spectra}
\tablehead{
\colhead{Source$^a$} &
\colhead{H$\alpha$ Flux$^b$} &
\colhead{\oi\,6300$^c$} &
\colhead{H$\alpha$} &
\colhead{\nii\,6584$^c$} &
\colhead{\sii\,6716$^c$} &
\colhead{\sii\,6731$^c$} &
\colhead{H$\alpha$ FWHM$^d$} &
\colhead{\sii\ FWHM$^d$}
}
\startdata
W23-01 & 46.1$\pm$0.9 & -- & 300 & 27.0$\pm$4.5 & 69.6$\pm$2.2 & 51.1$\pm$2.1 & 61.6$\pm$1.3 & 65.2$\pm$2.0 \\ 
{\bf W23-02} & 41.5$\pm$0.7 & 46.5$\pm$2.8 & 300 & 34.6$\pm$4.0 & 91.1$\pm$2.3 & 73.0$\pm$2.2 & 95.2$\pm$1.8 & 101.5$\pm$2.5 \\ 
W23-03 & 15.3$\pm$1.1 & -- & 300 & 41.8$\pm$17.0 & 130.0$\pm$12.8 & 69.4$\pm$11.4 & 222.6$\pm$17.3 & 212.4$\pm$21.7 \\ 
W23-06 & 22.5$\pm$0.6 & -- & 300 & 47.7$\pm$6.3 & 95.9$\pm$5.0 & 82.0$\pm$4.9 & 112.4$\pm$3.3 & 123.4$\pm$6.0 \\ 
W23-07 & 13.5$\pm$0.5 & -- & 300 & 66.9$\pm$9.3 & 128.2$\pm$6.2 & 99.6$\pm$5.9 & 142.9$\pm$6.0 & 90.2$\pm$4.2 \\ 
W23-08 & 37.3$\pm$0.8 & 45.2$\pm$7.8 & 300 & 49.3$\pm$5.5 & 115.0$\pm$2.7 & 86.9$\pm$2.5 & 78.7$\pm$2.0 & 75.1$\pm$1.7 \\ 
{\bf W23-10} & 230.0$\pm$1.7 & 57.9$\pm$1.4 & 300 & 53.7$\pm$1.8 & 155.2$\pm$1.3 & 104.9$\pm$1.2 & 92.4$\pm$0.8 & 96.1$\pm$0.8 \\ 
W23-11 & 19.4$\pm$0.2 & -- & 300 & 55.1$\pm$3.0 & 97.0$\pm$3.3 & 71.8$\pm$3.1 & 64.7$\pm$0.9 & 66.0$\pm$2.2 \\ 
{\bf W23-13} & 33.8$\pm$0.4 & 39.3$\pm$5.5 & 300 & 72.6$\pm$2.8 & 155.3$\pm$2.9 & 114.5$\pm$2.7 & 110.9$\pm$1.4 & 123.7$\pm$2.2 \\ 
{\bf W23-14} & 36.0$\pm$0.6 & 90.8$\pm$8.5 & 300 & 43.2$\pm$4.0 & 169.2$\pm$3.7 & 113.3$\pm$3.4 & 126.6$\pm$2.4 & 118.3$\pm$2.6 \\ 
W23-16 & 52.0$\pm$0.5 & -- & 300 & 34.3$\pm$2.2 & 57.2$\pm$1.0 & 39.9$\pm$0.9 & 59.7$\pm$0.6 & 56.6$\pm$0.9 \\ 
W23-18 & 2700.0$\pm$36.1 & 20.6$\pm$0.4 & 300 & 58.8$\pm$3.3 & 78.7$\pm$1.1 & 65.0$\pm$1.1 & 136.2$\pm$2.0 & 112.8$\pm$1.5 \\ 
{\bf W23-20} & 185.0$\pm$2.6 & 56.9$\pm$1.7 & 300 & 56.9$\pm$3.5 & 143.8$\pm$1.2 & 102.5$\pm$1.1 & 139.6$\pm$2.2 & 138.0$\pm$1.2 \\ 
W23-22 & 6780.0$\pm$42.8 & 6.5$\pm$0.5 & 300 & 43.7$\pm$1.5 & 46.9$\pm$0.5 & 38.2$\pm$0.5 & 101.9$\pm$0.7 & 111.3$\pm$1.2 \\ 
W23-23 & 740.0$\pm$6.2 & 36.5$\pm$0.7 & 300 & 39.9$\pm$2.0 & 77.0$\pm$1.3 & 62.4$\pm$1.3 & 78.7$\pm$0.7 & 103.7$\pm$1.7 \\ 
W23-25 & 266.0$\pm$3.5 & 22.8$\pm$0.8 & 300 & 62.6$\pm$3.2 & 105.5$\pm$1.5 & 75.6$\pm$1.4 & 167.9$\pm$2.5 & 157.3$\pm$2.1 \\ 
{\bf W23-28} & 175.0$\pm$2.1 & 95.7$\pm$1.8 & 300 & 47.5$\pm$3.0 & 174.9$\pm$2.2 & 125.0$\pm$2.0 & 110.5$\pm$1.5 & 124.1$\pm$1.5 \\ 
W23-31 & 71.2$\pm$0.7 & 66.6$\pm$1.3 & 300 & 38.0$\pm$2.5 & 139.5$\pm$1.2 & 98.6$\pm$1.1 & 88.6$\pm$1.0 & 88.1$\pm$0.7 \\ 
{\bf W23-35} & 55.2$\pm$0.9 & 36.2$\pm$3.9 & 300 & 49.4$\pm$4.1 & 133.2$\pm$2.9 & 98.4$\pm$2.7 & 114.2$\pm$2.2 & 151.1$\pm$3.2 \\ 
{\bf W23-36} & 11.9$\pm$0.6 & 67.1$\pm$14.3 & 300 & 59.2$\pm$12.8 & 140.4$\pm$9.0 & 107.9$\pm$8.5 & 112.2$\pm$6.5 & 114.3$\pm$7.1 \\ 
W23-37 & 17.4$\pm$0.2 & 49.0$\pm$7.5 & 300 & 69.8$\pm$2.3 & 108.8$\pm$2.6 & 76.0$\pm$2.4 & 70.5$\pm$0.7 & 63.3$\pm$1.5 \\ 
{\bf W23-38} & 74.2$\pm$1.3 & 56.2$\pm$7.0 & 300 & 24.6$\pm$4.2 & 117.7$\pm$2.6 & 78.4$\pm$2.4 & 97.4$\pm$1.9 & 145.4$\pm$3.3 \\ 
{\bf W23-39} & 108.0$\pm$1.2 & 36.9$\pm$1.9 & 300 & 77.8$\pm$2.7 & 122.5$\pm$1.9 & 88.6$\pm$1.8 & 92.4$\pm$1.1 & 82.7$\pm$1.3 \\ 
W23-40 & 136.0$\pm$1.7 & 27.1$\pm$1.9 & 300 & 49.0$\pm$3.0 & 103.7$\pm$1.3 & 73.5$\pm$1.2 & 93.1$\pm$1.3 & 74.2$\pm$0.9 \\ 
{\bf W23-41} & 693.0$\pm$1.7 & 38.1$\pm$0.4 & 300 & 46.3$\pm$0.6 & 117.7$\pm$0.4 & 83.1$\pm$0.4 & 91.9$\pm$0.3 & 84.7$\pm$0.3 \\ 
{\bf W23-42} & 188.0$\pm$2.0 & 59.4$\pm$1.0 & 300 & 49.5$\pm$2.6 & 126.2$\pm$4.4 & 123.5$\pm$4.4 & 95.4$\pm$1.2 & 107.1$\pm$3.4 \\ 
W23-44 & 138.0$\pm$1.1 & 50.2$\pm$1.4 & 300 & 38.3$\pm$2.0 & 115.9$\pm$1.6 & 80.9$\pm$1.5 & 65.1$\pm$0.6 & 79.3$\pm$1.1 \\ 
{\bf W23-45} & 104.0$\pm$1.0 & 49.3$\pm$1.9 & 300 & 53.7$\pm$2.3 & 107.3$\pm$1.2 & 91.2$\pm$1.2 & 161.4$\pm$1.7 & 167.8$\pm$1.7 \\ 
{\bf W23-47} & 81.5$\pm$0.9 & 54.5$\pm$1.3 & 300 & 39.0$\pm$2.7 & 125.2$\pm$1.8 & 86.1$\pm$1.7 & 94.0$\pm$1.2 & 93.5$\pm$1.3 \\ 
{\bf W23-48} & 64.8$\pm$0.8 & 51.9$\pm$2.8 & 300 & 72.2$\pm$2.9 & 153.2$\pm$2.0 & 103.7$\pm$1.8 & 125.3$\pm$1.6 & 131.2$\pm$1.7 \\ 
\hline
LBZ49$^e$ & 77.7$\pm$0.6 & 10.2$\pm$2.9 & 300 & 40.9$\pm$1.8 & 72.2$\pm$1.2 & 52.5$\pm$1.2 & 75.8$\pm$0.6 & 73.8$\pm$1.3 \\ 
LBZ67$^e$ & 182.0$\pm$1.3 & 15.1$\pm$0.9 & 300 & 45.3$\pm$1.7 & 77.0$\pm$0.6 & 56.4$\pm$0.6 & 64.0$\pm$0.5 & 69.3$\pm$0.5 \\ 
\enddata
\tablenotetext{a}{Objects in bold pass all three spectroscopic tests, and hence are almost certainly SNRs.}
\tablenotetext{b}{Flux in units of 10$^{-17}$ ergs cm$^{-2}$ s$^{-1}$}
\tablenotetext{c}{Ratio to H$\alpha$  flux where, by convention, H$\alpha$ is normalized to 300.}
\tablenotetext{d}{In units of  km s$^{-1}$.}
\tablenotetext{e}{ \cite{leonidaki13} candiates for which we obtained spectra.}
\label{snr_spectra}
\end{deluxetable}

\begin{deluxetable}{rrrrrrrrr}
\tabletypesize{\scriptsize}
\decimals
\tablecaption{NGC~4449 \hii\ region spectra}
\tablehead{
\colhead{Source} &
\colhead{H$\alpha$ Flux$^a$} &
\colhead{\oi\,6300$^b$} &
\colhead{H$\alpha$} &
\colhead{\nii\,6584$^b$} &
\colhead{\sii\,6716$^b$} &
\colhead{\sii\,6731$^b$} &
\colhead{H$\alpha$ FWHM$^c$} &
\colhead{\sii\ FWHM$^c$}
}
\startdata
H~II-01 & 20.4$\pm$0.5 & -- & 300 & 26.0$\pm$5.7 & 100.4$\pm$6.1 & 70.9$\pm$5.7 & 74.2$\pm$2.0 & 76.6$\pm$4.6 \\ 
H~II-02 & 14.9$\pm$0.3 & -- & 300 & 15.1$\pm$4.2 & 50.9$\pm$4.4 & 42.5$\pm$4.2 & 63.8$\pm$1.2 & 58.6$\pm$4.8 \\ 
H~II-03 & 30.8$\pm$0.3 & -- & 300 & 39.2$\pm$2.0 & 51.0$\pm$1.9 & 36.6$\pm$1.8 & 61.2$\pm$0.6 & 60.4$\pm$2.2 \\ 
H~II-04 & 117.0$\pm$0.5 & -- & 300 & 38.5$\pm$1.0 & 52.3$\pm$0.9 & 39.0$\pm$0.9 & 62.6$\pm$0.3 & 60.8$\pm$1.0 \\ 
H~II-05 & 13.0$\pm$0.3 & -- & 300 & 51.0$\pm$6.0 & 168.0$\pm$6.8 & 123.9$\pm$6.4 & 74.8$\pm$2.1 & 80.9$\pm$3.2 \\ 
H~II-06 & 49.1$\pm$0.4 & -- & 300 & 30.5$\pm$2.1 & 61.1$\pm$1.5 & 46.3$\pm$1.4 & 69.9$\pm$0.7 & 73.7$\pm$1.7 \\ 
H~II-07 & 42.3$\pm$0.3 & 19.0$\pm$3.7 & 300 & 49.0$\pm$1.9 & 79.4$\pm$1.7 & 57.6$\pm$1.6 & 76.6$\pm$0.7 & 75.9$\pm$1.6 \\ 
H~II-08 & 11.8$\pm$0.3 & -- & 300 & 37.6$\pm$5.2 & 96.9$\pm$6.8 & 54.9$\pm$6.1 & 64.4$\pm$1.6 & 60.8$\pm$4.4 \\ 
H~II-09 & 133.0$\pm$1.6 & 14.9$\pm$4.4 & 300 & 51.7$\pm$3.0 & 83.7$\pm$2.1 & 54.1$\pm$1.9 & 85.1$\pm$1.2 & 77.8$\pm$2.0 \\ 
H~II-10 & 963.0$\pm$12.6 & 4.9$\pm$0.3 & 300 & 37.7$\pm$3.2 & 36.4$\pm$0.5 & 27.9$\pm$0.5 & 76.6$\pm$1.1 & 52.7$\pm$0.7 \\ 
H~II-11 & 191.0$\pm$0.5 & -- & 300 & 25.3$\pm$0.7 & 30.0$\pm$0.4 & 20.1$\pm$0.4 & 63.7$\pm$0.2 & 56.8$\pm$0.8 \\ 
H~II-12 & 70.0$\pm$0.4 & 16.4$\pm$1.6 & 300 & 45.4$\pm$1.6 & 82.7$\pm$0.8 & 60.4$\pm$0.8 & 69.7$\pm$0.5 & 64.3$\pm$0.6 \\ 
H~II-13 & 63.7$\pm$0.2 & 9.7$\pm$2.0 & 300 & 27.1$\pm$0.8 & 44.1$\pm$0.7 & 30.9$\pm$0.6 & 61.9$\pm$0.2 & 58.2$\pm$0.9 \\ 
\enddata
\tablenotetext{a}{Flux in units of 10$^{-17}$ ergs cm$^{-2}$ s$^{-1}$}
\tablenotetext{b}{Ratio to H$\alpha$  flux where, by convention, H$\alpha$ is normalized to 300.}
\tablenotetext{c}{In units of km s$^{-1}$.}
\label{h2_spectra}
\end{deluxetable}

\begin{deluxetable}{lcccccccc}
\tabletypesize{\scriptsize}
\decimals
\tablecaption{Summary of Tests of SNR Candidates and H~II regions in NGC~4449}
\tablehead{
\colhead{Group} &
\colhead{Sample Size} &
\colhead{\sii:H$\alpha^a$} &
\colhead{\oi:H$\alpha^b$} &
\colhead{FWHM$^c$} &
\colhead{\sii:H$\alpha$ \& \oi:H$\alpha$} &
\colhead{\sii:H$\alpha$ \& FWHM} &
\colhead{\oi:H$\alpha$ \& FWHM} &
\colhead{All}
}
\startdata
SNRs & 30 & 25 & 20 & 22 & 19 & 20 & 15 & 15 \\ 
Morph(A) & 22 & 18 & 18 & 16 & 17 & 14 & 13 & 13 \\ 
Morph(B) & \phn8 & \phn7 & \phn2 & \phn6 & \phn2 & \phn6 & \phn2 & \phn2 \\ 
\hii\ regions & 13 & \phn3 & \phn0 & \phn0 & \phn0 & \phn0 & \phn0 & \phn0 \\ 
\enddata
\tablenotetext{^a}{\sii:H$\alpha>$ 0.5}
\tablenotetext{^b}{\oi:H$\alpha>$ 0.1}
\tablenotetext{^c}{FWHM $>$ 90 km s$^{-1}$}
\label{stats}
\end{deluxetable}

\end{document}